\documentclass[11pt]{article}
\pdfoutput=1

\usepackage{a4wide}
\usepackage[fleqn]{amsmath}
\usepackage{bbding}

\usepackage{pifont}
\usepackage{booktabs}
\usepackage{siunitx}

\usepackage{fourier} 
\usepackage[scaled=0.875]{helvet} 

\usepackage{tikz}
\usetikzlibrary{matrix}

\tikzset{
  pretableaumatrix/.style={
    ampersand replacement=\&,
    matrix of math nodes,
    outer sep=1mm,
    inner sep=0mm,
    anchor=center,
    row sep={between borders,-\pgflinewidth},
    column sep={between borders,-\pgflinewidth},
    dottedentry/.style={densely dotted},
    dashedentry/.style={densely dashed},
    spaceentry/.style={draw=none,execute at begin node=\null},
  },
  pretableaunode/.style={
    font=\small,
    draw=gray,
    sharp corners,
    rectangle,
    anchor=base,
    text height=3.75mm,
    text depth=1.25mm,
    minimum height=5mm,
    minimum width=5mm,
    inner sep=0mm,
    outer sep=0mm,
    doublewidth/.style={minimum width=10mm},
    footnotesize/.style={font=\footnotesize},
    scriptsize/.style={font=\scriptsize},
  },
  tableaumatrix/.style={
    pretableaumatrix,
    every node/.append style={
      pretableaunode,
    },
  },
  medtableaumatrix/.style={
    pretableaumatrix,
    every node/.append style={
      pretableaunode,
      font=\footnotesize,
      text height=2.75mm,
      text depth=.75mm,
      minimum height=3.5mm,
      minimum width=3.5mm
    },
  },
  smalltableaumatrix/.style={
    pretableaumatrix,
    every node/.append style={
      pretableaunode,
      font=\scriptsize,
      text height=1.85mm,
      text depth=.15mm,
      minimum height=2.5mm,
      minimum width=2.5mm,
    },
  },
  tinytableaumatrix/.style={
    pretableaumatrix,
    every node/.append style={
      pretableaunode,
      font=\tiny,
      text height=1.25mm,
      text depth=.15mm,
      minimum height=1.75mm,
      minimum width=1.75mm
    },
  },
  tableau/.style={
    baseline=-1.25mm,
    every matrix/.style={tableaumatrix},
  },
  medtableau/.style={
    baseline=-1.25mm,
    every matrix/.style={medtableaumatrix},
  },
  smalltableau/.style={
    baseline=-1.25mm,
    every matrix/.style={smalltableaumatrix},
  },
  preshapetableaumatrix/.style={
    pretableaumatrix,
    execute at end cell={\strut},
    every node/.append style={
      draw=black,
      anchor=base,
      inner sep=0mm,
      outer sep=0mm,
    },
    shadedentry/.style={fill=gray},
    darkshadedentry/.style={fill=darkgray},
  },
  medshapetableaumatrix/.style={
    preshapetableaumatrix,
    every node/.append style={
      text height=2.75mm,
      text depth=.75mm,
      minimum height=3.5mm,
      minimum width=3.5mm
    },
  },
  shapetableaumatrix/.style={
    ampersand replacement=\&,
    matrix of math nodes,
    outer sep=0mm,
    inner sep=0mm,
    anchor=base,
    row sep={between borders,-\pgflinewidth},
    column sep={between borders,-\pgflinewidth},
    execute at begin cell={\strut},
    every node/.append style={draw,anchor=base,text height=1mm,text depth=.5mm,minimum size=1.5mm,inner sep=0mm,outer sep=0mm},
  },
  shapetableau/.style={
    every matrix/.style={shapetableaumatrix},
  },
  topalign/.style={
    every matrix/.append style={name=maintableau,anchor=maintableau-1-1.base},
    baseline,
  },
}

\RequirePackage{amsmath,amssymb,amsxtra,amsthm}

\RequirePackage{mathrsfs}

\RequirePackage{setspace}
\RequirePackage{array}
\RequirePackage{booktabs}

\RequirePackage{colortbl}
\RequirePackage{xcolor}
\colorlet{titlerowcolor}{gray!15}

\usepackage[square,numbers]{natbib}

\definecolor{blue3}{RGB}{31,119,180}
\definecolor{red3}{RGB}{214,39,40}
\definecolor{orange3}{RGB}{255,127,14}
\definecolor{green3}{RGB}{44,160,44}

\usepackage{colortbl}
\definecolor{lightgreen}{cmyk}{0.2, 0, 0.2, 0.2}
\definecolor{lightgray}{cmyk}{0.1,0.2,0,0.1}
\definecolor{lightgray2}{cmyk}{0.1,0.1,0,0.1}

\usepackage{hyperref}
\hypersetup{colorlinks=true,linkcolor=teal,citecolor=orange3,urlcolor=green3,pdfencoding=auto}

\numberwithin{equation}{section}
\numberwithin{table}{section}
\numberwithin{figure}{section}

\author{
  \begin{minipage}{1.00\linewidth}
    \vspace{1cm}
    \begin{center}
      \begin{small}
        \textbf{Carlo Angelantonj$^{1,2}$, Ioannis Florakis$^3$, Giorgio Leone$^{1,2}$ and Diego Perugini$^1$}
        \\ \vspace{1cm}
        ${}^1$ {\em Dipartimento di Fisica, Universit\`a di Torino,  Via Pietro Giuria 1, I-10125 Torino}
        \\
        ${}^2$ {\em INFN Sezione di Torino, Via Pietro Giuria 1, I-10125 Torino}
        \\
        ${}^3$ {\em Department of Physics, University of Ioannina, GR-45110, Ioannina}
     \end{small}
    \end{center}
    \vspace{1cm}
  \end{minipage}
}

\date{}

\title{\vspace{3cm}
  \begin{huge} \textbf{Non-supersymmetric non-tachyonic heterotic vacua with reduced rank in various dimensions} 	
  \end{huge}
  \\ \vspace{.7cm}
}

\begin{document}

\begin{titlepage}
  \maketitle
  \thispagestyle{empty}

  \vspace{-14cm}
  \begin{flushright}
   \end{flushright}

  \vspace{11cm}

  \begin{center}
    \textsc{Abstract}\\
  \end{center}
We construct rigid non-supersymmetric heterotic vacua with reduced rank and no tachyons in six and four dimensions. These configurations are based on asymmetric orbifold compactifications which do not admit neutral deformation moduli and represent, to the best of our knowledge, the first instances of non-tachyonic non-supersymmetric heterotic vacua with reduced rank.  

\vfill

{\small
\begin{itemize}
\item[E-mail:] {\tt carlo.angelantonj@unito.it}
\\
{\tt iflorakis@uoi.gr}
\\
{\tt giorgio.leone@unito.it}
\\
{\tt diego.perugini@edu.unito.it}
\end{itemize}
}

\end{titlepage}

\setstretch{1.1}


{		\hypersetup{linkcolor=black}
		\tableofcontents	}

\section{Introduction}

Breaking supersymmetry in string theory unavoidably leads to instabilities. In the worst case scenario, the light spectrum is plagued by tachyonic scalars which clearly signal that the model does not even correspond to a classical vacuum. The spectrum of excitations  does not describe the correct particle content of the theory, and any attempt to compute quantum corrections is doomed. The fate of these configurations is hard to control since an off-shell potential is not available in the string theory description, and the condensation process is practically intractable, save for very few exceptions \cite{Antoniadis:1991kh, Antoniadis:1999gz, Kaidi:2020jla, Hellerman:2004qa}. 

Seemingly tachyon-free models are not difficult to build in perturbative string theory. These configurations are typically accompanied by a number of massless scalars which probe the classical moduli space. They parametrise marginal deformations of the two-dimensional sigma-model and allow one to tour the scalar manifold which includes regions where tachyons strike back. Clearly, in the absence of supersymmetry, quantum corrections are expected to lift the moduli space and generate an effective potential which could shield us from these dangerous regions. Although a complete analysis is still lacking, studies \cite{Ginsparg:1986wr, Fraiman:2023cpa} have nevertheless shown that the tachyon-free extrema of the scalar potential actually correspond to maxima or saddle points, so that one is unavoidably driven towards tachyonic domains. 

On the contrary, truly tachyon-free models, which are those that cannot be continuously deformed into classically unstable vacua, are hard to construct. In ten dimensions only three such cases exist corresponding to the celebrated $\text{SO}(16) \times \text{SO} (16)$ heterotic string \cite{Alvarez-Gaume:1986ghj, Dixon:1986iz}, the $0'\text{B}$ orientifold \cite{Sagnotti:1995ga, Sagnotti:1996qj} and the type IIB orientifold with USp(32) gauge group \cite{Sugimoto:1999tx}. In lower dimensions, 
ways to build classically stable orientifold vacua have been known since long time, and use suitable combinations of O-planes and D-branes, known as Brane Supersymmetry Breaking \cite{Antoniadis:1999xk, Aldazabal:1999jr, Angelantonj:1999jh, Angelantonj:1999ms, Angelantonj:2024iwi}, or non-trivial implementations of the Scherk-Schwarz mechanism \cite{Angelantonj:2006ut}. In the case of oriented closed strings, however, no such examples were known until very recently, when new non-supersymmetric non-tachyonic heterotic vacua were constructed in eight, six and four dimensions \cite{Baykara:2024tjr}, based on asymmetric (quasi-crystalline \cite{Harvey:1987da,Baykara:2024vss}) orbifolds.

Inspired by this work, and by a renewed interest \cite{ Font:2021uyw, Nakajima:2023zsh, DeFreitas:2024ztt} in CHL-like models \cite{Chaudhuri:1995fk, Chaudhuri:1995bf}, we present here new non-supersymmetric non-tachyonic heterotic vacua with reduced rank in six and four dimensions, which cannot be continuously deformed to tachyonic ones. They are based on orbifold compactifications of the $\text{SO}(16) \times \text{SO} (16)$ string involving the permutation of the two gauge group factors and suitable rotations and shifts acting along the right-moving and left-moving compact string coordinates, respectively. To the best of our knowledge, they are the first instances of genuinely stable non-supersymmetric CHL-like constructions in any dimension.

Although asymmetric orbifolds have a long history \cite{Narain:1986qm, Narain:1990mw},  they have been rarely studied in the literature aside from the pioneering work \cite{Ibanez:1987pj}, and some more recent attempts to build a heterotic vacuum with interesting phenomenology \cite{Beye:2013ola}, or study of the moduli space of string compactifications \cite{Acharya:2022shu, Gkountoumis:2023fym, Gkountoumis:2024dwc}. Their main attractiveness, however, lies in the fact that asymmetric actions are typically compatible only with rigid tori. As a result, the light spectrum involves very few moduli \cite{Bianchi:2012xz, Anastasopoulos:2009kj, Bianchi:1999uq, Baykara:2023plc, Baykara:2024tjr}, and provides a special incarnation  \cite{Condeescu:2012sp, Condeescu:2013yma} of non-geometric flux compactifications. The combination of permutation symmetry and asymmetric orbifolds, which project away K\"ahler, complex structure and Wilson line moduli, is precisely the key to building non-supersymmetric heterotic vacua of reduced rank which are both classically stable and rigid, in the sense that there are no 
neutral scalars which can be used to deform the classical vacuum.

The paper is organised as follows. In Section \ref{Sec:T4Z2} we focus on six dimensions and construct a simple example based on an asymmetric $\mathbb{Z}_2$ orbifold resulting in a rigid, chiral, tachyon-free vacuum with gauge group $\text{SO} (8) \times \text{SO} (16)$ and no dangerous neutral moduli. In Section \ref{Sec:T6Z2Z2}, we extend this construction down to four dimensions based on asymmetric $\mathbb{Z}_2 \times \mathbb{Z}_2$ orbifolds utilising both rotations and translations, and present three different non-supersymmetric stable vacua with rank fourteen and twenty-two, though non-chiral. In all cases, we display the full one-loop partition function, extract the massless spectrum and compute the corresponding vacuum energy. In Section \ref{Sec:deformations} we elaborate on the (im)possibility to deform these vacua by giving non-trivial vacuum expectation values (\emph{vev}'s) to adjoint\footnote{These massless scalars are not conventional Wilson lines, but are associated to the anti-diagonal combination of the ten-dimensional $\text{SO}(16)\times \text{SO} (16)$ generators which, thus, transform in the adjoint representation of the diagonal gauge group surviving the orbifold action. The nature of these scalars will be further clarified in the following sections.} and charged massless scalars.

\section{The asymmetric $T^4/\mathbb{Z}_2$ orbifold}
\label{Sec:T4Z2}

Let us start from the $\text{SO}(16) \times \text{SO} (16)$ heterotic string \cite{Alvarez-Gaume:1986ghj, Dixon:1986iz} compactified on a $T^4$ at the special point of $\text{SO} (8)$ symmetry enhancement, with background metric and $B$ field
\begin{equation}
g_{ij} = \begin{pmatrix} 2 & 0 & -1 & 0 \\ 0 & 2 & -1 & 0 \\ -1 & -1 & 2 & -1 \\ 0 & 0 & -1 & 2\end{pmatrix}\,, \qquad 
B_{ij} = \begin{pmatrix} 0 & 0 & -1 & 0 \\ 0 & 0 & -1 & 0 \\ 1 & 1 & 0 & -1 \\ 0 & 0 & 1 & 0 \end{pmatrix} \,.
\end{equation}
The partition function\footnote{Here and in the following we omit in ${\mathscr Z}$ the invariant integration over the fundamental domain of the modular group $\text{SL} (2,\mathbb{Z})$ and the contribution $\tau_2^{-d/2}\, (\eta (q) \, \bar \eta (\bar q))^{-d}$ associated to the $d$ transverse non compact world-sheet bosons, including their zero modes.}
\begin{equation}
\begin{split}
{\mathscr Z} =& \left[ \bar V_8 \, (O_{16} O_{16} + S_{16} S_{16} ) - \bar S_8 \, (O_{16} S_{16} + S_{16} O_{16} ) - \bar C_8 \, (V_{16} V_{16} + C_{16} C_{16} ) + \bar O_8 \, (V_{16} C_{16} + C_{16} V_{16} ) \right] 
\\
&\times \left( |O_8 |^2 + |V_8|^2 + |S_8 |^2 + |C_8 |^2 \right)\,,
\end{split}\label{6dtorus}
\end{equation}
is written in terms of $\text{SO} (2n)$ level-one characters, whose definition in terms of Dedekind and Jacobi theta functions and modular properties can be conveniently found in \cite{Angelantonj:2002ct}. Here, the anti-holomorphic characters are associated to the $\text{NS}_\pm$ and $\text{R}_\pm$ sectors of the world-sheet right-moving fermions, the SO(16) characters are associated to the left-moving gauge degrees of freedom, while the second line encodes the contribution of the compact bosons associated to the $T^4$ which, at this special point, factorises as a sesquilinear combination of left and right moving SO(8) characters. 

The NS vacuum from $\bar O_8$ is not level-matched, and therefore this compactification does not include tachyons. However, the massless spectrum contains, aside from the ubiquitous dilaton, sixteen neutral scalars, corresponding to the K\"ahler and complex structure moduli of the $T^4$, together with 64 Wilson lines which can then be used to deform the vacuum and connect it to tachyonic models. As anticipated in the Introduction, although quantum corrections typically lift the moduli space, a detailed study of the Coleman-Weinberg potential reveals that non tachyonic extrema only correspond to maxima or saddle points, and one is naturally driven towards unstable regions \cite{Ginsparg:1986wr, Fraiman:2023cpa}. 

For this reason, following \cite{Bianchi:2012xz, Anastasopoulos:2009kj, Bianchi:1999uq} and \cite{Baykara:2023plc} we mod-out the vacuum \eqref{6dtorus} by the asymmetric $\mathbb{Z}_2$ orbifold acting on the right-moving fields along the $T^4$ directions as
\begin{equation}
\gamma : \quad X^i_\text{R} (\tau -\sigma ) \to - X^i_\text{R} (\tau -\sigma )\,, \qquad \tilde\psi^i (\tau -\sigma  ) \to - \tilde\psi^i (\tau -\sigma  )\,.
\label{6dZ2action}
\end{equation}
This choice is compatible with modular invariance and particle interpretation of the one-loop vacuum energy only for our choice of factorised $T^4$, and does not require any action on the gauge degrees of freedom, since the phases generated by the $T$ modular transformation are cancelled in the holomorphic and anti-holomorphic sectors, independently. 
The vertex operators $\partial X^i \, (\bar\partial X^j + i p\cdot \tilde \psi\,  \tilde \psi^j ) e^{i p\cdot X}$ and $J^a\, (\bar\partial X^j + i p\cdot \tilde \psi\,  \tilde \psi^j ) e^{i p\cdot X}$ associated to the complex structure and K\"ahler moduli and Wilson lines, $J^a$ being an holomorphic $\text{SO} (16) \times \text{SO} (16)$ current, are odd with respect to the $\mathbb{Z}_2$ generator \eqref{6dZ2action} and are therefore projected away from the physical spectrum. This is clearly a consequence of the asymmetric action which is not compatible with a generic shape and size of the compactification torus. However, as we shall see, these states will come back from the twisted sector, unless we accompany the asymmetric rotation with an additional involution which acts non-trivially on the gauge degrees of freedom. 

The orbifold naturally breaks the SO(8) symmetries associated to the world-sheet fermions and the compact right-moving bosons down to $\text{SO} (4) \times \text{SO} (4)$, with
\begin{equation}
\bar V_8 = \bar V_4 \bar O_4 + \bar O_4 \bar V_4\,, \qquad \bar O_8 = \bar O_4 \bar O_4 + \bar V_4 \bar V_4 \,, \qquad
\bar S_8 = \bar C_4 \bar C_4 + \bar S_4 \bar S_4\,, \qquad \bar C_8 = \bar S_4 \bar C_4 + \bar C_4 \bar S_4 \,,
\label{so4so4}
\end{equation}
and the asymmetric $\mathbb{Z}_2$ rotation \eqref{6dZ2action} on the compact right-moving coordinates acts on the anti-holomorphic characters as
\begin{equation}
\begin{split}
\bar V_4 \bar O_4 + \bar O_4 \bar V_4 & \to \bar V_4 \bar O_4 - \bar O_4 \bar V_4\,,
\\
\bar O_4 \bar O_4 + \bar V_4 \bar V_4 & \to \bar O_4 \bar O_4 - \bar V_4 \bar V_4 \,,
\end{split}
\qquad
\begin{split}
\bar C_4 \bar C_4 + \bar S_4 \bar S_4 &\to \bar C_4 \bar C_4 - \bar S_4 \bar S_4 \,,
\\
\bar S_4 \bar C_4 + \bar C_4 \bar S_4 &\to \bar S_4 \bar C_4 - \bar C_4 \bar S_4 \,.
\end{split}\label{asymRot}
\end{equation}
The orbifold partition function then reads
\begin{equation}
{\mathscr Z} = \tfrac{1}{2} \sum_{h,g=0,1} {\mathscr Z} \big[ {\textstyle{h\atop g}}\big] \,,
\label{6dpart}
\end{equation}
where ${\mathscr Z} \big[ {\textstyle{0\atop 0}}\big]$ is simply given by the toroidal amplitude \eqref{6dtorus}, 
\begin{equation}
\begin{split}
{\mathscr Z} \big[ {\textstyle{0\atop 1}}\big] =&  \left[ (\bar V_4 \bar O_4 - \bar O_4 \bar V_4 ) (O_{16} O_{16}+S_{16} S_{16} ) 
+ (\bar O_4 \bar O_4 - \bar V_4 \bar V_4 ) (C_{16} V_{16}+V_{16} C_{16} )\right.
\\
&\qquad \left.
-(\bar C_4 \bar C_4 - \bar S_4 \bar S_4 ) (S_{16} O_{16}+O_{16} S_{16} )
-(\bar S_4 \bar C_4 - \bar C_4 \bar S_4 ) (V_{16} V_{16}+C_{16} C_{16} ) \right]
\\
&\times \left[(\bar O_4 \bar O_4 - \bar V_4 \bar V_4 ) O_8 + (\bar V_4 \bar O_4 - \bar O_4 \bar V_4 ) V_8 + (\bar C_4 \bar C_4 - \bar S_4 \bar S_4 ) S_8 + (\bar S_4 \bar C_4 - \bar C_4 \bar S_4 ) C_8 \right]\,,
\end{split}
\end{equation}
counts the untwisted states modulo their $\mathbb{Z}_2$ parity, while the amplitudes ${\mathscr Z} \big[ {\textstyle{1\atop 0}}\big]$ and ${\mathscr Z} \big[ {\textstyle{1\atop 1}}\big]$, defining the twisted sector, are obtained from ${\mathscr Z} \big[ {\textstyle{0\atop 1}}\big]$ upon the action of the $S$ and $TS$ modular transformations, respectively.

\begin{table}
\centering
\begin{tabular}{lllll} \toprule
{Sector} & {   } & {fields} & {   } & {$\text{SO} (16) \times \text{SO} (16) \times \text{SO} (8)$} 
\\
\midrule
{Untwisted}   & &   {$g_{\mu\nu}, B_{\mu\nu} , \phi$} & & {$(\textbf{1}, \textbf{1} , \textbf{1} )$}
 \\
 & &  {$A_\mu$} & & {$(\textbf{120}, \textbf{1}, \textbf{1})+(\textbf{1}, \textbf{120}, \textbf{1})+(\textbf{1},\textbf{1}, \textbf{28})$}
 \\
 & & {$\psi_\text{L}$} & & {$(\textbf{128}, \textbf{1}, \textbf{1} )+(\textbf{1}, \textbf{128}, \textbf{1} )$}
 \\
 & & {$\psi_\text{R}$} & & {$(\textbf{16},\textbf{16}, \textbf{1} )$}
 \\
 \midrule
 {Twisted}  & & {$\psi_\text{L}$} & & {$(\textbf{128}, \textbf{1}, \textbf{1} )+(\textbf{1}, \textbf{128}, \textbf{1} )$}
 \\
 & & {$\psi_\text{R}$} & & {$(\textbf{16},\textbf{16}, \textbf{1} )$}
 \\
 & & {$4\, \phi$} & & {$(\textbf{120}, \textbf{1}, \textbf{1})+(\textbf{1}, \textbf{120}, \textbf{1})+(\textbf{1},\textbf{1}, \textbf{28})$}
 \\
 & & {$4\, A_\mu$} & & {$(\textbf{1}, \textbf{1}, \textbf{1})$}
 \\
\bottomrule
\end{tabular}\\
\caption{The light spectrum of the six-dimensional $\mathbb{Z}_2$ orbifold associated to the partition function \eqref{6dpart}. The scalar field $\phi$ in the first line is the universal dilaton, while $\psi_\text{L}$ and $\psi_\text{R}$ denote right-handed (LHMW) and right-handed (RHMW) Majorana-Weyl  fermions, respectively.}
\label{6dspect}
\end{table}

This six-dimensional vacuum is non-tachyonic and the massless spectrum, collected in table \ref{6dspect}, is non-chiral. 
As anticipated,  this spectrum is nothing but that of the toroidal compactification of the ten-dimensional $\text{SO} (16) \times \text{SO} (16)$ theory. Indeed, the eight Abelian vectors, including the Cartan generators of SO(8), are nothing but the mixed components $g_{\mu i}$ and $B_{\mu i}$ while, among the twisted scalars there are sixteen neutral ones, \emph{i.e.} the four scalars associated to the four Cartan generators of SO(8), which are nothing but the deformation moduli of the $T^4$. These scalars, together with the Wilson lines of the $\text{SO} (16) \times \text{SO} (16)$ gauge group, can acquire a non-trivial \emph{vev} and take us away from the $ \text{SO} (16) \times \text{SO} (16)\times \text{SO}(8)$ point. Therefore, this model can be continuously connected to tachyonic vacua. 

In order to get a rigid vacuum, we can combine the asymmetric rotation \eqref{6dZ2action} with a (modular invariant) action on the gauge degrees of freedom. The simplest and most interesting choice is to use the permutation $R$ of the two $\text{SO} (16)$ groups. Taking into account that only the $\bar V_8 (O_{16} O_{16} + S_{16} S_{16})$ and $\bar C_8 (V_{16} V_{16} + C_{16} C_{16})$ contributions admit a non-trivial action of $R$, the full partition function now reads
\begin{equation}
{\mathscr Z}_\text{6d} = \tfrac{1}{2} \sum_{h,g=0,1} {\mathscr Z}_\text{6d} \big[ {\textstyle{h\atop g}}\big] \,,
\label{6dCHL}
\end{equation}
with
\begin{equation}
\begin{split}
{\mathscr Z}_\text{6d} \big[ {\textstyle{0\atop 0}}\big] =& \left[ (\bar V_4 \bar O_4 + \bar O_4 \bar V_4 ) (O_{16} O_{16}+S_{16} S_{16} ) 
+ (\bar O_4 \bar O_4 + \bar V_4 \bar V_4 ) (C_{16} V_{16}+V_{16} C_{16} )\right.
\\
&\qquad \left.
-(\bar C_4 \bar C_4 + \bar S_4 \bar S_4 ) (S_{16} O_{16}+O_{16} S_{16} )
-(\bar S_4 \bar C_4 + \bar C_4 \bar S_4 ) (V_{16} V_{16}+C_{16} C_{16} ) \right]
\\
&\times \left[(\bar O_4 \bar O_4 + \bar V_4 \bar V_4 ) O_8 + (\bar V_4 \bar O_4 + \bar O_4 \bar V_4 ) V_8 + (\bar C_4 \bar C_4 + \bar S_4 \bar S_4 ) S_8 + (\bar S_4 \bar C_4 + \bar C_4 \bar S_4 ) C_8 \right]\,,
\end{split}\label{CHL6d00}
\end{equation}
\begin{equation}
\begin{split}
{\mathscr Z}_\text{6d} \big[ {\textstyle{0\atop 1}}\big]=& \left[ (\bar V_4 \bar O_4 -\bar O_4 \bar V_4 ) ( O_{16}(q^2) + S_{16} (q^2) )-  ( \bar S_4 \bar C_4 - \bar C_4 \bar S_4  ) ( V_{16}(q^2) + C_{16} (q^2)  ) \right]
\\
&\times \left[(\bar O_4 \bar O_4 - \bar V_4 \bar V_4 ) O_8 + (\bar V_4 \bar O_4 - \bar O_4 \bar V_4 ) V_8 + (\bar C_4 \bar C_4 - \bar S_4 \bar S_4 ) S_8 + (\bar S_4 \bar C_4 - \bar C_4 \bar S_4 ) C_8 \right]\,,
\end{split}\label{CHL6d01}
\end{equation}
\begin{equation}
\begin{split}
{\mathscr Z}_\text{6d} \big[ {\textstyle{1\atop 0}}\big] =&   \left[ (\bar O_4 \bar C_4 + \bar V_4 \bar S_4 -\bar S_4 \bar V_4 -\bar C_4 \bar O_4 )  O_{16}(\sqrt{q})
  +  (\bar O_4 \bar S_4 + \bar V_4 \bar C_4 -\bar C_4 \bar V_4 -\bar S_4 \bar O_4  ) S_{16} (\sqrt{q}) )\right]
\\
&\times   \left[ ( \bar O_4 \bar S_4 + \bar V_4 \bar C_4  ) O_8 +  (\bar V_4 \bar S_4 + \bar O_4 \bar C_4  ) \bar{V}_8 +  ( \bar S_4 \bar O_4 + \bar C_4 \bar V_4  ) S_8 +  ( \bar S_4 \bar V_4 + \bar C_4 \bar O_4  ) C_8\right]\,,
\end{split}\label{CHL6d10}
\end{equation}
and
\begin{equation}
\begin{split}
{\mathscr Z}_\text{6d} \big[ {\textstyle{1\atop 1}}\big] =&  \left[ ( \bar O_4 \bar C_4 - \bar V_4 \bar S_4 +\bar S_4 \bar V_4 - \bar C_4 \bar O_4  )  \hat O_{16}(-\sqrt{q}) 
+  (\bar O_4 \bar S_4 - \bar V_4 \bar C_4 + \bar C_4 \bar V_4 - \bar S_4 \bar O_4  ) \hat S_{16} (-\sqrt{q}) \right] 
\\
&\times \left[ ( \bar V_4 \bar C_4 - \bar O_4 \bar S_4   ) O_8 +  (\bar O_4 \bar C_4   - \bar V_4 \bar S_4  ) V_8 + ( \bar S_4 \bar O_4 - \bar C_4 \bar V_4  ) S_8 +  (\bar C_4 \bar O_4 - \bar S_4 \bar V_4  ) C_8\right]\,.
\end{split}\label{CHL6d11}
\end{equation}
Notice that ${\mathscr Z}_\text{6d} \big[ {\textstyle{0\atop 0}}\big]$ is a rewriting of eq. \eqref{6dtorus}, where the anti-holomorphic SO(8) characters have been conveniently decomposed as in \eqref{so4so4}, and is manifestly modular invariant. The twisted amplitudes ${\mathscr Z}_\text{6d} \big[ {\textstyle{1\atop 0}}\big]$ and ${\mathscr Z}_\text{6d} \big[ {\textstyle{1\atop 1}}\big]$ are, as usual, obtained from ${\mathscr Z}_\text{6d} \big[ {\textstyle{0\atop 1}}\big]$ via the action of $S$ and $TS$ modular transformations, so that also the complete partition function \eqref{6dCHL} is $\text{SL} (2,\mathbb{Z} )$ invariant. To lighten the notation, in ${\mathscr Z}_\text{6d} \big[ {\textstyle{1\atop 1}}\big]$ we have introduced the \emph{hatted} characters $\hat O_{16}(-\sqrt{q}) = e^{i\pi /3}\, O_{16} (-\sqrt{q}) $ and $\hat S_{16}(-\sqrt{q}) = e^{i\pi /3}\, S_{16} (-\sqrt{q}) $, which absorb an overall phase generated by the $T$ modular transformation. Crucially, the $q$-expansion of the partition function only involves integer coefficients.   

\begin{table}
\centering
\begin{tabular}{lllll} \toprule
{Sector} & {   } & {fields} & {   } & {$\text{SO} (16)_2 \times \text{SO} (8)_1$} 
\\
\midrule
{Untwisted}   & &   {$g_{\mu\nu}, B_{\mu\nu} , \phi$} & & {$(\textbf{1}, \textbf{1} )$}
 \\
 & &  {$A_\mu$} & & {$(\textbf{120}, \textbf{1})+(\textbf{1}, \textbf{28})$}
 \\
 & & {$\psi_\text{L}+\psi_\text{R}$} & & {$(\textbf{128}, \textbf{1} )$}
 \\
 & & {$\psi_\text{R}$} & & {$(\textbf{136}, \textbf{1} )$}
 \\
 & & {$4\phi + \psi_\text{L}$} & & {$(\textbf{120}, \textbf{1} )$}
 \\
 \midrule
 {Twisted}  & & {$4\phi + \psi_\text{L}$} & & {$(\textbf{120} ,\textbf{1})+(\textbf{1}, \textbf{8}_v + \textbf{8}_s + \textbf{8}_c )$}
\\
 & & {$4\phi + \psi_\text{R}$} & & {$(\textbf{128} ,\textbf{1})$}
 \\
\bottomrule
\end{tabular}\\
\caption{The light spectrum of the rigid six-dimensional $\mathbb{Z}_2$ orbifold associated to the partition function \eqref{6dCHL}. The suffix $k$ in the gauge group indicates that the associated Ka$\check{\text{c}}$-Moody algebra is realised at level $k$, \emph{i.e.}  $\text{SO} (16)$ has level $k=2$ while for $\text{SO} (8)$ $k=1$. The scalar field $\phi$ in the first line is the universal dilaton, while $\psi_\text{L}$ and $\psi_\text{R}$ denote right-handed (LHMW) and right-handed (RHMW) Majorana-Weyl  fermions, respectively.}
\label{6dspectrum}
\end{table}

In this case, the six-dimensional massless spectrum is chiral and is reported in table \ref{6dspectrum}. Notice that, as expected, the K\"ahler and complex structure moduli together with the Wilson lines are absent, which reflects the fact that this vacuum cannot be deformed. Untwisted massless scalars,  associated to the term $\frac{1}{2} \, \bar O_4 \bar V_4 \, O_{16} O_{16}$ in ${\mathscr Z}_\text{6d} \big[ {\textstyle{0\atop 0}}\big]$,  transform in the adjoint representation of the gauge group and are a signature of vacua where the permutation of the gauge factors is accompanied by the asymmetric rotation. Indeed, they are also present in the supersymmetric vacua of \cite{Baykara:2023plc} where a similar orbifold is constructed for the $\text{E}_8 \times \text{E}_8$ heterotic string. 

To better understand their origin, and the fact that these scalars are \emph{not bona fide} Wilson lines of the surviving gauge group, we note that the two SO(16) currents $J^a_{1,2}$ of the parent theory should be conveniently decomposed into the eigenstates $J^a_\pm \sim J^a_1 \pm J^a_2$ of the permutation operator $R$ with eigenvalues $\pm 1$. Only the diagonal generators $J_+^a$ form a subgroup, while the $J_-^a$ do not close into a subalgebra, and transform in the adjoint representation of the diagonal subgroup, $[J_+^a , J_-^b ] = i f^{ab}{}_c\, J_-^c$.  As a result, the vertex operators 
\begin{equation}
J^a_+\, (\bar\partial X^\mu + i p\cdot \tilde \psi \, \tilde \psi^\mu ) e^{ip\cdot X}\qquad \text{and} \qquad
J^a_-\, (\bar\partial X^i + i p\cdot \tilde \psi \, \tilde \psi^i ) e^{ip\cdot X}\,,
\label{evencombs}
\end{equation}
invariant under the action of the full orbifold, describe gauge bosons for the physical gauge group $ \text{SO} (16)_2 $ and the (untwisted) scalars $\phi^i$ associated to the anti-diagonal generators, but still transforming in the adjoint representation. This description of the spectrum is indeed compatible with the partition function, since 
\begin{equation}
\begin{split}
{\mathscr Z}_\text{6d} \big[ {\textstyle{0\atop 1}}\big] &\supset  (\bar V_4 \bar O_4 -\bar O_4 \bar V_4 ) O_{16}(q^2) \simeq (\bar V_4 \bar O_4 -\bar O_4 \bar V_4 ) (q^{-1}+ 120 \, q )  
\\
&\simeq  (\bar V_4 \bar O_4 -\bar O_4 \bar V_4 ) (q^{-1} + (120-120) \, q^0 + 120 \, q )\,,
\end{split}
\end{equation}
where the coefficient of $q^0$ counts the number of $J^a_\pm$ currents graded by their $\mathbb{Z}_2$ eigenvalue. 

This vacuum is rigid in the sense that it does not contain neutral scalars associated to the geometric moduli and Wilson lines. Whether or not charged scalars can acquire a \emph{vev} and break the gauge group is an interesting question, whose unambiguous answer would, however, require a complete knowledge of the scalar potential which is difficult to reconstruct. We defer a discussion on these (im)possible deformations to Section \ref{Sec:deformations}.

It is straightforward to show that this chiral spectrum has vanishing irreducible gauge and gravitational anomalies, while the residual anomaly polynomial factorises as
\begin{equation}
I_8 = \tfrac{1}{4} \left( 2 \text{tr} \, R^2 - \tfrac{2}{2} \text{tr} \, F_{16}^2 - \tfrac{1}{2} \text{tr} \, F_{8}^2 \right) \left( \tfrac{1}{2} \text{tr} \, F_{16}^2 - \tfrac{1}{4} \text{tr} \, F_{8}^2 \right) \,,
\end{equation}
and is cancelled by a six-dimensional version of the Green-Schwarz mechanism \cite{Green:1984sg, Sagnotti:1992qw}. Notice that this anomaly polynomial takes the standard form \cite{Schellekens:1986yi, Schellekens:1986xh} for six-dimensional heterotic vacua and the coefficients of the $\text{tr} \, F_{16}^2$ and $\text{tr} \, F_{8}^2 $ reflect the fact that the SO(16) gauge group is realised at level $k=2$, while the SO(8) group at $k=1$. It would be interesting to study non-perturbative anomalies along the lines of \cite{Basile:2023zng}. 

This model, defined by the vacuum amplitudes \eqref{CHL6d00}, \eqref{CHL6d01}, \eqref{CHL6d10} and \eqref{CHL6d11} is, to the best of our knowledge, the first instance of a non-supersymmetric CHL-like construction which is free of tachyonic excitations and cannot be continuously deformed. It represents the six-dimensional analogue of the ten-dimensional $\text{SO} (16) \times \text{SO} (16)$ heterotic string, but with reduced rank gauge group realised at level two. 

\begin{figure}
	\centering
	\includegraphics[width=9cm]{"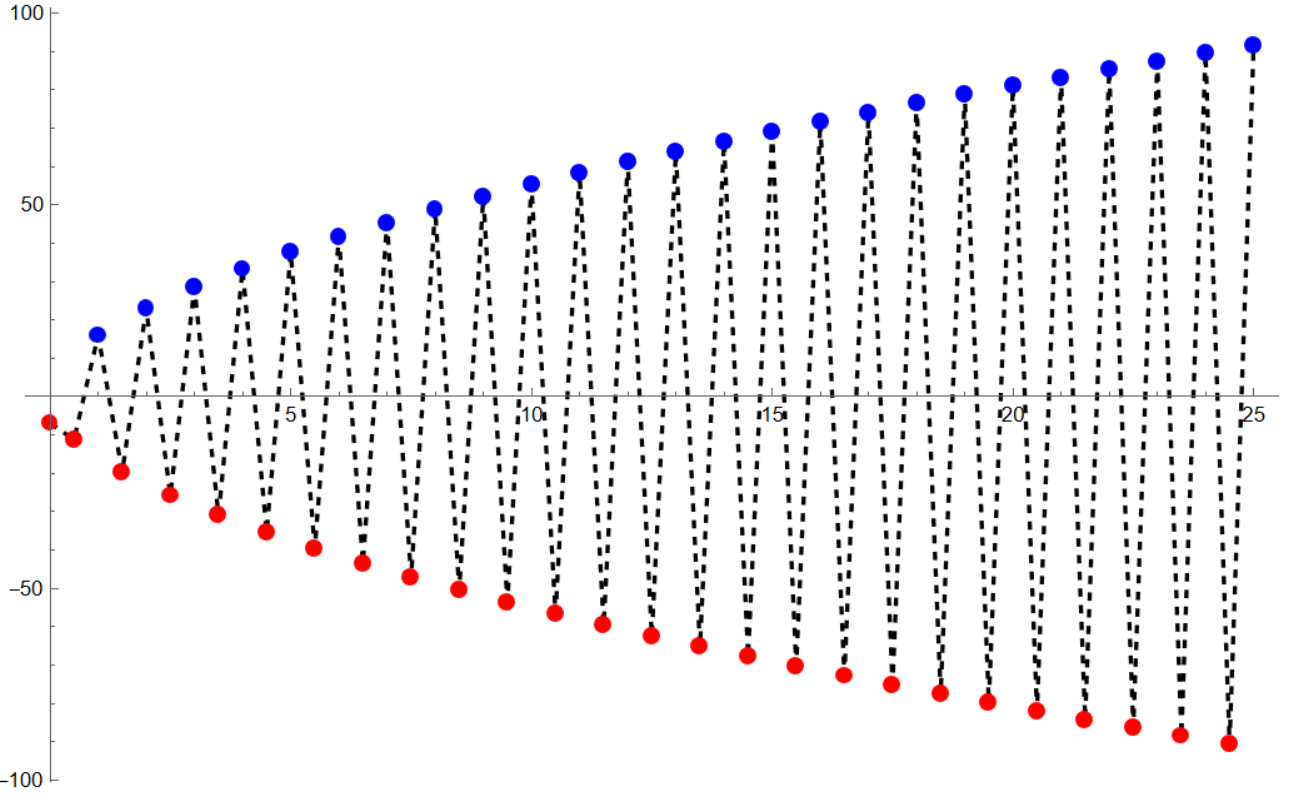"}
	\caption{The {\em signed} logarithm of the net number of degrees of freedom at each mass level for the six-dimensional non-tachyonic CHL-like vacuum with gauge group $\text{SO}(8)_1 \times \text{SO}(16)_2$ highlighting the presence of misaligned supersymmetry. Positive (negative) contributions are ascribed to the excess of bosonic (fermionic) states.}
	\label{6dmisaligned}
\end{figure}

The classical stability of this vacuum is associated to the presence of misaligned supersymmetry \cite{Kutasov:1990sv, Dienes:1994np,  Angelantonj:2010ic, Cribiori:2020sct, Angelantonj:2023egh, Leone:2023qfd} in the tower of massive string excitations. Indeed, following \cite{Angelantonj:2023egh} it is straightforward to compute the exact sector averaged sum and show that the effective central charge vanishes, $C_\text{eff} =0$, so that the overall (signed) spectrum does not exhibit an exponential growth signalling the onset of a Hagedorn phase transition at some critical temperature. This is reflected in the oscillating pattern of the net number of degrees of freedom represented in figure \ref{6dmisaligned}. Clearly, being non-supersymmetric, this model has a non-vanishing vacuum energy
\begin{equation}
\Omega = - \frac{1}{(2\pi \sqrt{\alpha '})^6}\int_{\mathscr F} \frac{d\tau_1 d\tau_2}{2\tau_2^4}\, \frac{{\mathscr Z}_\text{6d} (\tau_1 , \tau_2 )}{(\eta \bar\eta )^4}\simeq 3.31 \times 10^{-3} \, \alpha'{}^{-3}\,,
\end{equation}
which is positive and  induces a one-loop runaway potential for the dilaton.

This asymmetric $\mathbb{Z}_2$ orbifold can be evidently used to mod-out any theory which respects $\gamma\, R$, as for instance  the $[\text{E}_7 \times \text{SU} (2)]^2$ non-supersymmetric tachyonic heterotic string \cite{Dixon:1986iz} compactified on the SO(8) maximal torus, 
\begin{equation}
\begin{split}
{\mathscr Z}_t \big[{\textstyle{0 \atop 0}}\big] =& \left[ \bar V_8  \, ( \chi_o \xi_o \chi_o \xi_o + \chi_v \xi_v \chi_v \xi_v ) - \bar S_8 \, (\chi_o \xi_o \chi_v \xi_v + \chi_v \xi_v \chi_o \xi_o ) \right.
\\
&\qquad \left. - \bar C_8 \, (\chi_o \xi_v \chi_v \xi_o +\chi_v \xi_o \chi_o \xi_v ) + \bar O_8 \, (\chi_o \xi_v \chi_o \xi_v + \chi_v \xi_o \chi_v \xi_o ) \right]
\\
&\times  \left( |O_8 |^2 + |V_8 |^2 + |S_8 |^2 + |C_8 |^2 \right)\,.
\end{split}
\end{equation}
Here the characters $\chi_o$ and $\chi_v$ ($\xi_o$ and $\xi_v$) are associated to the conjugacy classes of the adjoint and fundamental representations of SU(2) ($\text{E}_7$). As in the previous case, we decompose the anti-holomorphic SO(8) characters into product of $\text{SO} (4) \times \text{SO} (4)$ ones \eqref{so4so4}, so that the asymmetric rotation has the natural action \eqref{asymRot} and the untwisted projected amplitude reads
\begin{equation}
\begin{split}
{\mathscr Z}_t \big[{\textstyle{0 \atop 1}}\big] =& \left[ \left (\bar V_4 \bar O_4 -\bar O_4 \bar V_4 \right ) \left ( \chi_o \xi_o + \chi_v \xi_v \right )  (q^2) 
     + \left ( \bar O_4 \bar O_4 -\bar V_4 \bar V_4 \right ) \left ( \chi_o \xi_v + \chi_v \xi_o \right )(q^2)\right]
\\
&\times  \left[ ( \bar O_4 \bar O_4 -\bar V_4 \bar V_4  ) O_8 +  (\bar V_4 \bar O_4 -\bar O_4 \bar V_4  ) V_8 +  ( \bar C_4 \bar C_4 -\bar S_4 \bar S_4  ) S_8 +  ( \bar S_4 \bar C_4 - \bar C_4 \bar S_4  ) C_8\right]\,.
\end{split}
\end{equation}
As usual, the twisted amplitudes ${\mathscr Z}_t \big[{\textstyle{1 \atop 0}}\big]$ and ${\mathscr Z}_t \big[{\textstyle{1 \atop 1}}\big]$ can be obtained from ${\mathscr Z}_t \big[{\textstyle{0 \atop 1}}\big]$ via the modular transformations $S$ and $TS$, respectively. The light spectrum includes, in the untwisted sector, the metric, the Kalb-Ramond field and the dilaton, gauge bosons for the gauge group $\left ( \text{E}_7 \times\text{SU}(2) \right)_2 \times \text{SO}(8)_1$, a tachyon in the $(\boldsymbol{1},\boldsymbol{2}, \boldsymbol{1})$ representation and two Majorana fermions in the $( \boldsymbol{56}, \boldsymbol{2}, \boldsymbol{1})$ representation. The twisted sector gives  four scalars in the $(\boldsymbol{1},\boldsymbol{3}, \boldsymbol{1})+(\boldsymbol{133},\boldsymbol{1}, \boldsymbol{1})+(\boldsymbol{1},\boldsymbol{1}, \boldsymbol{8}_v+\boldsymbol{8}_s +\boldsymbol{8}_c)$ representations and one extra Majorana fermion in the $(\boldsymbol{56},\boldsymbol{2}, \boldsymbol{1})$ representation. Finally, we get also in this case the signature adjoint scalars associated to the internal components of the odd combination of gauge bosons, as in eq. \eqref{evencombs}. 
Again, the K\"ahler and  complex structure moduli, as well as the Wilson lines are absent and thus this tachyonic spectrum is rigid.  Being non-chiral the anomaly polynomial vanishes identically, while the vacuum energy is divergent because of the presence of tachyons.

\section{The asymmetric $T^6/\mathbb{Z}_2\times \mathbb{Z}_2$ orbifold}
\label{Sec:T6Z2Z2}

The generalisation to four dimensions is, in principle,  straightforward. The naive choice would be to perform an asymmetric $\mathbb{Z}_2\times \mathbb{Z}_2$ orbifold, where each $\mathbb{Z}_2$ is a copy of the six-dimensional one, though acting along different directions,
\begin{equation}
\gamma_1 = (-,+,-)_\text{R} \, \times \, R\,, \qquad \gamma_2 = (+,-,-)_\text{R} \,\times \, R\,.
\label{4dz2z2}
\end{equation}
Indeed, at first sight, the Wilson lines $J^a\, (\bar\partial X^j + i p\cdot \tilde \psi\,  \tilde \psi^j ) e^{i p\cdot X}$ and the moduli $\partial X^i \, (\bar\partial X^j + i p\cdot \tilde \psi\,  \tilde \psi^j ) e^{i p\cdot X}$ of the compactification torus are both odd under the $\mathbb{Z}_2\times \mathbb{Z}_2$ action. However, a more attentive analysis shows that the element $\gamma_1 \gamma_2$ acts as a simple right-moving rotation of four coordinates, and leaves the left moving sector completely untouched. This element reproduces (the four-dimensional reduction of) the partition function \eqref{6dtorus}, and thus, per se, corresponds to a simple toroidal compactification of the heterotic string defined by eq. \eqref{6dCHL}, where (some of) the deformation moduli and Wilson lines come back from the twisted sector. To overcome this problem, and make sure that the deformation moduli from the $\gamma_1 \gamma_2$ twisted sector become massive, we accompany the right-moving rotations with left-moving (asymmetric) shifts. Modular invariance and the rules of free-fermionic construction \cite{Antoniadis:1986rn} (see \cite{Florakis:2024ubz} for a recent review) require that the asymmetric shift acts on four coordinates. One can then make three inequivalent choices for the orbifold generators
\begin{align}
 \mathbb{Z}_2 \times \mathbb{Z}_2 &:\qquad  \gamma_1 = (-,+,-)_\text{R} \, \times (1,\delta , \delta )_\text{L} \, \times R\,,
\qquad 
\gamma_2 = (+,-,-)_\text{R} \, \times (\delta,1,\delta  )_\text{L} \, \times R\,, \label{4d1}
\\
(\mathbb{Z}_2 \times \mathbb{Z}_2)^\prime &:\qquad  \gamma_1^{\prime} = (-,+,-)_\text{R} \, \times (1,\delta , \delta )_\text{L} \, \times R\,, 
\qquad 
\gamma_2^{\prime} = (+,-,-)_\text{R} \, \times (1,1 , 1 )_\text{L} \, \times R\,, \label{4d2}
\\
(\mathbb{Z}_2 \times \mathbb{Z}_2 )^{\prime\prime} &: \qquad \gamma_1^{\prime\prime} = (-,+,-)_\text{R} \, \times (1,\delta , \delta )_\text{L} \,,
\qquad \qquad \!
\gamma_2^{\prime\prime} = (+,-,-)_\text{R} \, \times (\delta,1,\delta )_\text{L} \,.  \label{4d3}
\end{align}
As usual, $R$ denotes the permutation of the two $\text{SO} (16)$ groups and therefore only the first and second choice yield a gauge group of reduced rank. The entries in the brackets refer to the three pairs of compact coordinates, each parametrising a $T^2$. A minus sign indicates a $\pi$ rotation on the right-moving coordinates of the associated two-plane, and similarly $\delta$ induces an order-two shift along the pair of left-moving coordinates. Clearly, this asymmetric action is only compatible with a special $T^6$ which we take to be the maximal torus of SO(12).

To better understand the action of rotation and shifts on the compact coordinates, it is convenient to resort to a description in terms of free fermions, so that $X^i_\text{L}$ and $X^i_\text{R}$ are described by  the real auxiliary fermions $ y^i , \omega^i$ and $\bar y^i , \bar\omega^i$, respectively.
Since rotations and shifts act on the bosonic coordinates as
\begin{equation}
X^i_\text{R} \to - X^i_\text{R}\,, \qquad 
X^i_\text{L} \to X^i_\text{L} + \pi\,,
\end{equation}
the fermionisation map
\begin{equation}
\bar y^i \sim \sin ( X^i_\text{R})\,, \qquad \bar\omega^i \sim \cos ( X^i_\text{R} )\,, 
\end{equation}
yields the following action on the free fermions
\begin{equation}
(\bar y^i , \bar\omega^i ) \to (-\bar y^i , +\bar\omega^i ) \,,
\qquad
(y^i , \omega^i ) \to (- y^i , -\omega^i ) \,.
\end{equation}
Moreover, the orbifold generators break the right-moving SO(12) as
\begin{equation}
\text{SO} (12)_\text{R} \to \left[ \text{SO} (2) \times \text{SO}(2) \right]^3_\text{R}\,, \label{RMbreak}
\end{equation}
while the left-moving SO(12) is broken down to 
\begin{equation}
\text{SO} (12)_\text{L} \to \left[ \text{SO} (4) \times \text{SO}(4) \times \text{SO} (4) \right]_\text{L} \label{LMbreak13}
\end{equation}
for the first and third choice, while
\begin{equation}
\text{SO} (12)_\text{L} \to \left[ \text{SO} (4) \times \text{SO}(8) \right]_\text{L} \label{LMbreak2}
\end{equation}
for the second one, since the shifts act simultaneously on the $3,4,5,6$ directions. 

In eq. \eqref{RMbreak} the first (second) SO(2) factor is associated to the pair $\bar y^{1,2}$ ($\bar \omega^{1,2}$) and similarly for the other SO(2)'s. In eq. \eqref{LMbreak13} the $k$-th SO(4) factor is associated to the four fermions $y^{2k-1,2k}, \,\omega^{2k-1,2k}$, respectively, while in eq. \eqref{LMbreak2} the SO(4) is associated to the four fermions $y^{1,2},\, \omega^{1,2}$ and SO(8) to the eight fermions $y^{3,4,5,6}, \omega^{3,4,5,6}$. Finally, in the RNS sector, the ten-dimen\-sional little group SO(8) is broken to $\text{SO} (2)^4$, parametrised by $\bar\chi^\mu$, for the non-compact directions, and $\bar\chi^i$,
for the compact ones. Putting everything together, the action of the three asymmetric orbifolds is specified in tables \ref{T4dorbifold1}, \ref{T4dorbifold2} and \ref{T4dorbifold3}, and reflects into
\begin{equation}
\begin{split}
(O_2, V_2 , S_2 , C_2 ) &\to (O_2 , -V_2 , i\, S_2 , -i\, C_2)\,, 
\\
(O_4 , V_4 , S_4 , C_4 ) &\to (O_4 , -V_4 , - S_4 , C_4)\,,
\\
(O_8 , V_8 , S_8 , C_8 ) &\to (O_8 , -V_8 , S_8 , -C_8)\,,
\end{split}
\end{equation}
for the corresponding SO(2), SO(4) and SO(8) characters.

 \begin{table}
\centering
\begin{tabular}{SSSSSSSSSSSSS} \toprule
{}  &   {$\bar\chi^{1,2}$} & {$\bar\chi^{3,4}$} & {$\bar\chi^{5,6}$} & {$y^{1,2}\,,\, \omega^{1,2}$}  &  {$y^{3,4}\,,\, \omega^{3,4}$} &  {$y^{5,6}\,,\, \omega^{5,6}$} & {$\bar y^{1,2}$} & {$\bar\omega^{1,2}$} & {$\bar y^{3,4}$} & {$\bar\omega^{3,4}$} & {$\bar y^{5,6}$} & {$\bar\omega^{5,6}$}
 \\
 \midrule
 {$\gamma_1$} & \ding{51} & \ding{55} & \ding{51} & \ding{55} & \ding{51} & \ding{51} & \ding{51} & \ding{55} & \ding{55} & \ding{55} & \ding{51} & \ding{55} 
 \\
{$\gamma_2$} & \ding{55}  & \ding{51} & \ding{51} & \ding{51} & \ding{55} & \ding{51}  & \ding{55} & \ding{55} & \ding{51} & \ding{55} & \ding{51} & \ding{55} 
\\
{$\gamma_1\gamma_2$} & \ding{51}   & \ding{51} & \ding{55} & \ding{51} & \ding{51} &\ding{55} & \ding{51}  & \ding{55} & \ding{51} & \ding{55} & \ding{55}  & \ding{55} 
\\
\bottomrule
\end{tabular}\\
\caption{The action of the asymmetric $\mathbb{Z}_2 \times \mathbb{Z}_2$ orbifold of eq. \eqref{4d1} on the left-moving and right-moving fermions. The symbol \ding{51} (\ding{55}) clearly indicates that the element $\gamma_a$ does (not) act on the corresponding fermions. $\gamma_1$ and $\gamma_2$ also act on the gauge degrees of freedom exchanging the two SO(16) factors.}
\label{T4dorbifold1}
\end{table}

\begin{table}
\centering
\begin{tabular}{SSSSSSSSSSSS} \toprule
{}  &   {$\bar\chi^{1,2}$} & {$\bar\chi^{3,4}$} & {$\bar\chi^{5,6}$} & {$(y\,,\, \omega)^{1,2}$}  &  {$(y\,,\, \omega)^{3,4,5,6}$} & {$\bar y^{1,2}$} & {$\bar\omega^{1,2}$} & {$\bar y^{3,4}$} & {$\bar\omega^{3,4}$} & {$\bar y^{5,6}$} & {$\bar\omega^{5,6}$}
 \\
 \midrule
 {$\gamma_1$} & \ding{51} & \ding{55} & \ding{51} & \ding{55} & \ding{51} & \ding{51} & \ding{55} & \ding{55} & \ding{55} & \ding{51} & \ding{55}
 \\
{$\gamma_2$} & \ding{55}  & \ding{51} & \ding{51} & \ding{55} & \ding{55} & \ding{55} & \ding{55} & \ding{51} & \ding{55} & \ding{51} & \ding{55} 
\\
{$\gamma_1\gamma_2$} & \ding{51}   & \ding{51} & \ding{55} & \ding{55} & \ding{51} & \ding{51}  & \ding{55} & \ding{51} & \ding{55} & \ding{55}  & \ding{55}
\\
\bottomrule
\end{tabular}\\
\caption{The action of the asymmetric $(\mathbb{Z}_2 \times \mathbb{Z}_2)^\prime$ orbifold of eq. \eqref{4d2} on the left-moving and right-moving fermions. The symbol \ding{51} (\ding{55}) clearly indicates that the element $\gamma_a$ does (not) act on the corresponding fermions. $\gamma_1$ and $\gamma_2$ also act on the gauge degrees of freedom exchanging the two SO(16) factors.}
\label{T4dorbifold2}
\end{table}

 \begin{table}
\centering
\begin{tabular}{SSSSSSSSSSSSS} \toprule
{}  &   {$\bar\chi^{1,2}$} & {$\bar\chi^{3,4}$} & {$\bar\chi^{5,6}$} & {$(y\,,\, \omega)^{1,2}$}  &  {$(y\,,\, \omega)^{3,4}$} &  {$(y\,,\, \omega)^{5,6}$} & {$\bar y^{1,2}$} & {$\bar\omega^{1,2}$} & {$\bar y^{3,4}$} & {$\bar\omega^{3,4}$} & {$\bar y^{5,6}$} & {$\bar\omega^{5,6}$}
 \\
 \midrule
 {$\gamma_1$} & \ding{51} & \ding{55} & \ding{51} & \ding{55} & \ding{51} & \ding{51} & \ding{51} & \ding{55} & \ding{55} & \ding{55} & \ding{51} & \ding{55} 
 \\
{$\gamma_2$} & \ding{55}  & \ding{51} & \ding{51} & \ding{51} & \ding{55} & \ding{51}  & \ding{55} & \ding{55} & \ding{51} & \ding{55} & \ding{51} & \ding{55} 
\\
{$\gamma_1\gamma_2$} & \ding{51}   & \ding{51} & \ding{55} & \ding{51} & \ding{51} &\ding{55} & \ding{51}  & \ding{55} & \ding{51} & \ding{55} & \ding{55}  & \ding{55} 
\\
\bottomrule
\end{tabular}\\
\caption{The action of the asymmetric $(\mathbb{Z}_2 \times \mathbb{Z}_2)^{\prime\prime}$ orbifold of eq. \eqref{4d3} on the left-moving and right-moving fermions. The symbol \ding{51} (\ding{55}) clearly indicates that the element $\gamma_a$ does (not) act on the corresponding fermions. Notice that now, $\gamma_1$ and $\gamma_2$ do not act on the gauge degrees of freedom.}
\label{T4dorbifold3}
\end{table}


We have now all the ingredients to build the new four-dimensional vacua, starting from the toroidal compactification on the maximal torus of SO(12)
\begin{equation}
\begin{split}
{\mathscr Z} &= \left[ \bar V_8 \, (O_{16}O_{16}+S_{16}S_{16}) - \bar S_8  \, (O_{16}S_{16}+S_{16}O_{16}) - \bar C_8  \, (V_{16}V_{16}+C_{16}C_{16}) + \bar O_8  \, (V_{16}C_{16}+C_{16}V_{16}) \right]
\\
&\times \left[ |O_{12}|^2 + |V_{12}|^2 + |S_{12}|^2 + |C_{12}|^2 \right]\,,
\end{split}
\end{equation}
and implementing the orbifolds \eqref{4d1}, \eqref{4d2} or \eqref{4d3} as indicated in tables \ref{T4dorbifold1}, \ref{T4dorbifold2}, \ref{T4dorbifold3}. The associated massless spectra are collected in tables \ref{4dspectrum1}, \ref{4dspectrum2} and \ref{4dspectrum3}, respectively. The main feature shared by  all three spectra is the absence of neutral scalars associated to the deformations of the $T^6$ and to the Wilson lines, thus reflecting once more the rigidity of these constructions. These non-tachyonic vacua are the four-dimensional counterparts of the $\text{SO} (16)\times \text{SO} (16)$ model \cite{Alvarez-Gaume:1986ghj, Dixon:1986iz}.  Although in the case of maximal rank gauge group similar constructions, based on quasi-crystalline orbifolds, have been discussed in \cite{Baykara:2024tjr},  the $\mathbb{Z}_2 \times \mathbb{Z}_2$ and $(\mathbb{Z}_2 \times \mathbb{Z}_2)^\prime$
orbifolds represent the first instances of truly non-tachyonic four-dimensional constructions with reduced rank. Again, the common feature of the rank-reduced spectra is the presence of untwisted massless scalars transforming in the adjoint representation of the diagonal $\text{SO}(16)_2$, which are not conventional Wilson lines and, as we shall discuss in Section \ref{Sec:deformations}, do not correspond to truly marginal deformations. 

\begin{table}
\centering
\begin{tabular}{lllll} \toprule
{Sector} & {   } & {fields} & {   } & {$\text{SO} (16)_2 \times \text{SO} (4)_1\times \text{SO} (4)_1\times \text{SO} (4)_1$} 
\\
\midrule
{Untwisted}   & &   {$g_{\mu\nu}, B_{\mu\nu} , \phi$} & & {$(\textbf{1}, \textbf{1}, \textbf{1}, \textbf{1} )$}
 \\
 & &  {$A_\mu$} & & {$(\textbf{120}, \textbf{1}, \textbf{1}, \textbf{1})+(\textbf{1}, \textbf{6}, \textbf{1}, \textbf{1})+(\textbf{1}, \textbf{1}, \textbf{6}, \textbf{1})+(\textbf{1}, \textbf{1}, \textbf{1}, \textbf{6})$}
 \\
 & & {$\psi_\text{D}$} & & {$2\, (\textbf{128}, \textbf{1} , \textbf{1} , \textbf{1})+(\textbf{136}, \textbf{1} , \textbf{1}, \textbf{1} )+(\textbf{120}, \textbf{1} , \textbf{1} , \textbf{1})$}
 \\
 & & {$2\, \phi$} & & {$ (\textbf{120}, \textbf{1}, \textbf{1}, \textbf{1} )+(\textbf{1}, \textbf{4}, \textbf{4}, \textbf{1} )+(\textbf{1}, \textbf{4}, \textbf{1}, \textbf{4} )+(\textbf{1}, \textbf{1}, \textbf{4}, \textbf{4} )$}
 \\
 \midrule
 {$\gamma_1$ twisted}  & & {$2\phi + \psi_\text{D}$} & & {$(\textbf{1},\textbf{1}, \textbf{2}_s +\textbf{2}_c , \textbf{2}_s+\textbf{2}_c )$}
\\
 \midrule
 {$\gamma_2$ twisted}  & & {$2\phi + \psi_\text{D}$} & & {$(\textbf{1},\textbf{2}_s +\textbf{2}_c , \textbf{1},  \textbf{2}_s+\textbf{2}_c )$}
\\
 \midrule
 {$\gamma_1\gamma_2$ twisted}  & & {$2\phi $} & & {$(\textbf{1}, \textbf{2}_s +\textbf{2}_c , \textbf{2}_s+\textbf{2}_c , \textbf{4})$}
 \\
\bottomrule
\end{tabular}\\
\caption{The light spectrum of the rigid four-dimensional $\mathbb{Z}_2\times \mathbb{Z}_2$ orbifold generated by \eqref{4d1}. The suffix $k$ in the gauge group indicates that the associated Ka$\check{\text{c}}$-Moody algebra is realised at level $k$, \emph{i.e.}  $\text{SO} (16)$ has level $k=2$ while all $\text{SO} (4)$'s  have $k=1$. The scalar field $\phi$ in the first line is the universal dilaton, while $\psi_\text{D}$ denotes a Dirac  fermion.}
\label{4dspectrum1}
\end{table}

\begin{table}
\centering
\begin{tabular}{lllll} \toprule
{Sector} & {   } & {fields} & {   } & {$\text{SO} (16)_2 \times \text{SO} (4)_1\times \text{SO} (8)_1$} 
\\
\midrule
{Untwisted}   & &   {$g_{\mu\nu}, B_{\mu\nu} , \phi$} & & {$(\textbf{1}, \textbf{1}, \textbf{1} )$}
 \\
 & &  {$A_\mu$} & & {$(\textbf{120}, \textbf{1}, \textbf{1})+(\textbf{1}, \textbf{6}, \textbf{1})+(\textbf{1}, \textbf{1}, \textbf{28})$}
 \\
 & & {$\psi_\text{D}$} & & {$2\, (\textbf{128}, \textbf{1} , \textbf{1} )+(\textbf{136}, \textbf{1} , \textbf{1} )+(\textbf{120}, \textbf{1} , \textbf{1} )$}
 \\
 & & {$2\, \phi$} & & {$(\textbf{120}, \textbf{1}, \textbf{1} )+(\textbf{1}, \textbf{4}, \textbf{8}_v )$}
 \\
 \midrule
 {$\gamma_1$ twisted}  & & {$2\phi + \psi_\text{D}$} & & {$(\textbf{1},\textbf{1}, \textbf{8}_s + \textbf{8}_c )$}
\\
 \midrule
 {$\gamma_2$ twisted}  & & {$2\phi + \psi_\text{D}$} & & {$(\textbf{120},\textbf{1},\textbf{1}) + (\textbf{128}, \textbf{1},\textbf{1}) + (\textbf{1}, \textbf{4},\textbf{1}) + (\textbf{1}, \textbf{1}, \textbf{8}_v )$}
\\
 \midrule
 {$\gamma_1\gamma_2$ twisted}  & & {$2\phi $} & & {$(\textbf{1},\textbf{4}, \textbf{8}_s + \textbf{8}_c )$}
 \\
\bottomrule
\end{tabular}\\
\caption{The light spectrum of the rigid four-dimensional $(\mathbb{Z}_2\times \mathbb{Z}_2)^\prime$ orbifold generated by \eqref{4d2}. The suffix $k$ in the gauge group indicates that the associated Ka$\check{\text{c}}$-Moody algebra is realised at level $k$, \emph{i.e.}  $\text{SO} (16)$ has level $k=2$ while both $\text{SO} (4)$ and $\text{SO} (8)$ have $k=1$. The scalar field $\phi$ in the first line is the universal dilaton, while $\psi_\text{D}$ denotes a Dirac  fermion.}
\label{4dspectrum2}
\end{table}

\begin{table}
\centering
\begin{tabular}{lllll} \toprule
{Sector} & {   } & {fields} & {   } & {$\text{SO} (16) \times \text{SO} (16) \times \text{SO} (4) \times \text{SO} (4) \times \text{SO} (4)$} 
\\
\midrule
{Untwisted}   & &   {$g_{\mu\nu}, B_{\mu\nu} , \phi$} & & {$(\textbf{1},\textbf{1}, \textbf{1}, \textbf{1}, \textbf{1} )$}
 \\
 & &  {$A_\mu$} & & {$(\textbf{120}, \textbf{1},\textbf{1}, \textbf{1}, \textbf{1})+(\textbf{1}, \textbf{120},\textbf{1}, \textbf{1}, \textbf{1})$}
 \\
 & & & & {$ +
 (\textbf{1}, \textbf{1}, \textbf{6}, \textbf{1}, \textbf{1})+(\textbf{1}, \textbf{1}, \textbf{1}, \textbf{6}, \textbf{1})+(\textbf{1},\textbf{1},  \textbf{1}, \textbf{1}, \textbf{6})$}
 \\
 & & {$\psi_\text{D}$} & & {$ (\textbf{128}, \textbf{1} ,\textbf{1} , \textbf{1} , \textbf{1})+(\textbf{1}, \textbf{128} ,\textbf{1} , \textbf{1}, \textbf{1} )+(\textbf{16}, \textbf{16} , \textbf{1} ,\textbf{1} , \textbf{1})$}
 \\
 & & {$2\, \phi$} & & {$(\textbf{1},\textbf{1}, \textbf{4}, \textbf{4}, \textbf{1} )+(\textbf{1}, \textbf{1},\textbf{4}, \textbf{1}, \textbf{4} )+(\textbf{1}, \textbf{1},\textbf{1}, \textbf{4}, \textbf{4} )$}
 \\
 \midrule
 {$\gamma_1$ twisted}  & & {$2\phi$} & & {$(\textbf{1}, \textbf{1},  \textbf{4}, \textbf{2}_s +\textbf{2}_c , \textbf{2}_s+\textbf{2}_c )$}
\\
 \midrule
 {$\gamma_2$ twisted}  & & {$2\phi$} & & {$(\textbf{1}, \textbf{1}, \textbf{2}_s +\textbf{2}_c , , \textbf{4}, \textbf{2}_s+\textbf{2}_c )$}
\\
 \midrule
 {$\gamma_1\gamma_2$ twisted}  & & {$2\phi $} & & {$(\textbf{1}, \textbf{1}, \textbf{2}_s +\textbf{2}_c , \textbf{2}_s+\textbf{2}_c , \textbf{4})$}
 \\
\bottomrule
\end{tabular}\\
\caption{The light spectrum of the rigid four-dimensional $(\mathbb{Z}_2\times \mathbb{Z}_2)^{\prime\prime}$ orbifold generated by \eqref{4d3}.  All gauge factors are realised in terms of Ka$\check{\text{c}}$-Moody algebras at level one. The scalar field $\phi$ in the first line is the universal dilaton, while $\psi_\text{D}$ denotes a Dirac  fermion.}
\label{4dspectrum3}
\end{table}

For completeness, we report the associated partition functions written in terms of Dedekind eta and Jacobi theta functions\footnote{Notice that for the definition of the Jacobi theta functions we adopt the conventions used by the free fermions community (see, for instance, \cite{Florakis:2024ubz})
\begin{equation}
\theta \big[{\textstyle{a\atop b}}\big] (z|\tau )= \sum_{m\in\mathbb{Z}} q^{\frac{1}{2} (m-a/2)^2} \, e^{2 \pi i (m-a/2 )(z-b/2 )}\,,
\end{equation}
where the periodic/antiperiodic boundary conditions along the two cycles of the world-sheet torus are associated to characteristics equal to 1 or 0. In the following we shall not display the arguments $z=0$ and $\tau$ of the theta functions, unless it is required by the construction.}
\begin{equation}
{\mathscr A}_{(\ell )} =\frac{1}{2^3}\, \frac{1}{(\eta \, \bar\eta)^2}\sum_{h , g =0,1}\, \sum_{h_1 , g_1 =0,1} \, \sum_{h_2 , g_2  =0,1} \bar{\Phi} \bigg[ {\textstyle{h\ h_1\ h_2 \atop g\ g_1 \ g_2}}\bigg] \, \Lambda_{(\ell)} \bigg[ {\textstyle{h_1\ h_2 \atop g_1 \ g_2}}\bigg]\, \Pi_{(\ell)} \bigg[ {\textstyle{h\ h_1\ h_2 \atop g\ g_1 \ g_2}}\bigg]\, \,.
\end{equation}
where $\ell =1,2,3$ refers to the orbifolds \eqref{4d1}, \eqref{4d2} and \eqref{4d3},  respectively. Here $(h,g)$ build the $\text{SO} (16) \times \text{SO} (16)$ vacuum from the $\text{E}_8 \times \text{E}_8$ heterotic string, while $(h_1,g_1)$ and $(h_2,g_2)$ implement the different $\mathbb{Z}_2 \times \mathbb{Z}_2$ orbifolds. The $\bar \Phi$ function 
\begin{equation}
\bar{\Phi} \bigg[ {\textstyle{h\ h_1\ h_2 \atop g\ g_1 \ g_2}}\bigg] = \frac{1}{2}\sum_{a,b=0,1} \frac{\bar\theta \big[ {\textstyle{a\atop b}}\big] \, \bar\theta \big[ {\textstyle{a+h_1\atop b+g_1}}\big] \bar\theta \big[ {\textstyle{a+h_2\atop b+g_2}}\big] \, \bar\theta \big[ {\textstyle{a-h_1-h_2\atop b-g_1-g_2}}\big]}{\bar\eta^4}\, (-1)^{a+b+ab+ga+hb+hg}\,,
\end{equation} 
encodes the universal contribution of the RNS fermions. The function $\Lambda_{(\ell)}$ is associated to the compactification lattice, and
\begin{equation}
\begin{split}
\Lambda_{(\ell)} \bigg[ {\textstyle{h_1\ h_2 \atop g_1 \ g_2}}\bigg] &= \frac{1}{2}\sum_{\gamma,\delta=0,1} \frac{\theta \big[{\textstyle{\gamma+\alpha h_2 \atop  \delta +\alpha g_2 }}\big]^2 \, \theta \big[{\textstyle{\gamma +h_1  \atop  \delta+g_1}}\big]^2\, \theta \big[{\textstyle{\gamma -h_1-\alpha h_2 \atop  \delta-g_1-\alpha g_2}}\big]^2}{\eta^6}\,  \frac{\bar\theta \big[{\textstyle{\gamma +h_1\atop  \delta+g_1}}\big] \, \bar\theta \big[{\textstyle{\gamma +h_2\atop  \delta+g_2}}\big] \, \bar\theta \big[{\textstyle{\gamma-h_1-h_2 \atop  \delta-g_1-g_2}}\big] \,\bar\theta \big[{\textstyle{\gamma  \atop  \delta}}\big]^3}{\bar\eta^6}
\,,
\end{split}
\end{equation}
where $\alpha =1$ for $\ell=1,3$, while $\alpha=0$ for $\ell=2$. Finally, the $\Pi_{(\ell)} $ functions refers to the gauge degrees of freedom, and for $\ell=3$
\begin{equation}
\Pi_{(3)} \bigg[ {\textstyle{h\ h_1\ h_2 \atop g\ g_1 \ g_2}}\bigg] =\frac{1}{4} \sum_{k,l=0,1}  \sum_{\rho,\sigma =0,1}  \frac{\theta\big[{\textstyle{k\atop l}}\big]^8}{\eta^8}\,\frac{\theta\big[{\textstyle{\rho\atop \sigma}}\big]^8}{\eta^8}\, (-1)^{g (k+\rho ) + h (l+\sigma )}
\end{equation}
for any choice of $h_{1,2}$ and $g_{1,2}$, while for $\ell=1,2$ 
\begin{equation}
\Pi_{(1,2)} \bigg[ {\textstyle{h\ 0 \ 0 \atop g\ 0 \ 0}}\bigg] = \Pi_{(1,2)} \bigg[ {\textstyle{h\ 0 \ 0 \atop g\ 1 \ 1}}\bigg] = \Pi_{(3)} \bigg[ {\textstyle{h\ h_1\ h_2 \atop g\ g_1 \ g_2}}\bigg] \,,
\end{equation}
and
\begin{equation}
\Pi_{(1,2)} \bigg[ {\textstyle{h\ 0 \ 0 \atop g\ 1 \ 0}}\bigg] = \Pi_{(1,2)} \bigg[ {\textstyle{h\ 0 \ 0 \atop g\ 0 \ 1}}\bigg]  = \frac{1}{2} \sum_{k,l=0,1}\left(\frac{\theta\big[{\textstyle{k\atop l}}\big] (2\tau) }{\eta (2\tau)} \right)^8\, (-1)^{h l} \,,
\end{equation}
all other cases being obtained by suitable $S$ and $T$ modular transformations. Notice that although these $\mathbb{Z}_2 \times \mathbb{Z}_2$ orbifolds admit the possibility to turn on  discrete torsion \cite{Vafa:1994rv}, it does not play any role since, in the independent modular orbit, the lattice contribution vanishes identically for any $\ell=1,2,3$, namely
\begin{equation}
\Lambda_{(\ell )} \bigg[ {\textstyle{1\ 0 \atop 0 \ 1}}\bigg] \equiv 0\,,
\end{equation}
together with its modular transformations.

The one loop vacuum energies 
\begin{equation}
\Omega_{(\ell )} = - \frac{1}{(2\pi \sqrt{\alpha '})^4}\int_{\mathscr F} \frac{d\tau_1 d\tau_2}{2\tau_2^3}\, {\mathscr A}_{(\ell )} (\tau_1 , \tau_2 )
\end{equation}
are all finite and positive, with
\begin{equation}
\Omega_{(1)} \simeq 3.94 \times 10^{-2}\, {\alpha^{\prime}}^{-2}\,,
\qquad
\Omega_{(2)} \simeq 5.85 \times 10^{-2}\, {\alpha^{\prime}}^{-2}\,,
\qquad
\Omega_{(3)} \simeq 7.55 \times 10^{-2}\, {\alpha^{\prime}}^{-2}\,,
\end{equation}
and thus induce a non-trivial runaway potential for the dilaton. Again, the finiteness of the free energies implies the presence of misaligned supersymmetry in the whole string spectrum \cite{Kutasov:1990sv, Dienes:1994np,  Angelantonj:2010ic, Cribiori:2020sct, Angelantonj:2023egh, Leone:2023qfd}, which signals the vanishing of the sector averaged sums \cite{Dienes:1994np, Cribiori:2020sct, Angelantonj:2023egh}.

\section{Discussion on deformations}\label{Sec:deformations}

A natural question about the vacua built in the previous sections is whether or not one can use the neutral and charged scalars present in the spectrum to deform the sigma model. The answer to this question is complicated by the highly asymmetric nature of the orbifold and by the fact that supersymmetry is no longer present to protect the moduli space. As a result, the validity of the two-step process of first building the model from string theory and then studying its deformations at the level of the effective field theory, typically followed in supersymmetric constructions (based on symmetric orbifolds), is no longer obvious. Instead, one should approach the construction of the vacuum and its deformations within a consistent string-theory setup, and the string construction should stand on its own both before and after the charged scalars acquire a \emph{vev}. Although, in principle, there is no conceptual obstruction in describing the process of Higgsing the gauge group from a fully-fledged sigma model framework, the actual study of these marginal deformations is still a daunting task. 

In order to get a glimpse of what is doable or not in this situation, let us study the possible marginal deformations along the neutral directions in the Cartan subalgebra for the untwisted adjoint scalars of eq. \eqref{evencombs}, which are always present in our vacua with reduced rank. Their vertex operator 
\begin{equation}
\bar\partial X^i\, (J^a_1 - J^a_2 )
\end{equation}
shows that these states are also charged with respect to the gauge group associated to the compactification torus. Indeed, if $e^{i p_\text{L}\cdot X_\text{L}}$ is a charged current building the adjoint representation of the right-moving lattice gauge group, \emph{i.e.} SO(8) in six dimensions, one has
\begin{equation}
\left[ e^{i p_\text{L}\cdot X_\text{L}} \,, \, \bar\partial X^i\, (J^a_1 - J^a_2 ) \right] = i p_\text{L}^i \, e^{i p_\text{L}\cdot X_\text{L}} \, (J^a_1 - J^a_2 ) \,.
\end{equation}
As a result, an arbitrary \emph{vev} for the neutral scalars \eqref{evencombs} automatically implies the breaking of the lattice gauge group.  

Another way of seeing this is to introduce the deformation already at the level of the toroidally compactified theory, before the orbifold even acts. The presence of Wilson lines  boosts the left and right moving momenta as 
\begin{equation}
\begin{split}
p_{I\, \text{L}} &= \sqrt{\frac{\alpha'}{2}} \left( m_i +\frac{1}{\alpha '} (g_{ij} + B_{ij} )n^j - Q^a Y^a_i + \frac{1}{2} Y_i^a\, Y^a_j n^j \,, \, \sqrt{\frac{2}{\alpha '}} \left(Q^a + Y^a_j  n^j \right) \right)\,,
\\
p_{i\, \text{R}} &= \sqrt{\frac{\alpha'}{2}} \left( m_i - \frac{1}{\alpha '}(g_{ij} - B_{ij} )n^j - Q^a  Y^a_i + \frac{1}{2} Y_i^a\, Y^a_j  n^j \right)\,,
\end{split}
\end{equation}
where, in the case at hand, $g_{ij}$ and $B_{ij}$ are the Cartan and adjacency matrices of SO(8) (in six dimensions) or SO(12) (in four dimensions). Again, the presence of Wilson lines not only breaks the Kac-Moody gauge group, but also the gauge group originating from the compact coordinates at the enhancement point, since the condition $p_\text{R}^2 =0$ and $m_i n^i = 1$ from level matching, is only possible for vanishing, or at best discrete values of $Y$.

Continuous Wilson lines disrupt the independence of the left and right moving zero modes of the compact coordinates, and no longer define proper chiral CFT's.
As a result, moving away from the rational point is incompatible with our asymmetric orbifold. Indeed, an asymmetric $\mathbb{Z}_2$ rotation along $d$ right-moving coordinates acts non trivially on the oscillators and sets $p_{i\, \text{R}}$ to zero. At a generic point in moduli space, this also requires $p_{i\, \text{L}} = 0$, so that
\begin{equation}
{\mathscr Z} \big[ {\textstyle{0 \atop 1}}\big] = {\mathscr C}\big[ {\textstyle{0 \atop 1}}\big] \, \left(\frac{\bar\eta}{\bar\theta \big[ {\textstyle{1 \atop 0}}\big]}\right)^{d/2}\, \frac{1}{\eta^d}\,,
\end{equation}
where ${\mathscr C}\big[ {\textstyle{0 \atop 1}}\big] $ encodes the contribution of all remaining world-sheet degrees of freedom, which are not relevant for the present discussion.  Although, at first sight there appears to be no immediate problem with ${\mathscr Z} \big[ {\textstyle{0 \atop 1}}\big]$, its modular transformations 
\begin{equation}
\begin{split}
{\mathscr Z} \big[ {\textstyle{1 \atop 0}}\big] &= {\mathscr C}\big[ {\textstyle{1 \atop 0}}\big] \, \left(\frac{\bar\eta}{\bar\theta \big[ {\textstyle{0 \atop 1}}\big]}\right)^{d/2}\, \frac{1}{(-i \tau)^{d/2}\, \eta^d}\,,
\\
{\mathscr Z} \big[ {\textstyle{1 \atop 1}}\big] &= {\mathscr C}\big[ {\textstyle{1 \atop 1}}\big] \, \left(\frac{\bar\eta}{\bar\theta \big[ {\textstyle{0 \atop 0}}\big]}\right)^{d/2}\, \frac{1}{(-i \tau-i)^{d/2}\, \eta^d}\, e^{-i \pi d/8}\,,
\end{split}
\end{equation}
defining the twisted sector, clearly clash with a consistent particle interpretation. As a result, higher-loop modular invariance is lost, rendering the construction inconsistent. What saves an asymmetric orbifold is that, at special points of moduli space, and \emph{only} there, the vanishing of $p_\text{R}$ does not automatically imply $p_\text{L} =0$ but, instead, left-moving Kaluza-Klein/winding states dress the contribution $\eta^{-d}$ of the oscillators, remove the powers of $\tau$ in the twisted sectors, and yield a consistent partition function enjoying multi-loop modular invariance. 

All this strongly suggests that the adjoint scalars of eq. \eqref{evencombs} cannot be cavalierly used to continuously deform our vacua. As for the remaining charged scalars, it is very likely that an arbitrary \emph{vev} would similarly spoil the consistency of the construction. It is interesting to note that a similar situation occurs in orientifold vacua based on a $\mathbb{Z}_4$ orbifold with \emph{Brane Supersymmetry Breaking} \cite{Angelantonj:2024iwi}. In this case, the O5 planes localised on the $\mathbb{Z}_4$ fixed points carry a twisted RR charge which, for consistency, must be \emph{locally} cancelled by fractional D5-branes sitting on the same fixed points. Untwisted and twisted  RR tadpole cancellation then implies that the geometry of branes supporting a $\text{USp} (4)^8$ gauge group is rigid, despite the presence of massless scalars in various bifundamental representations. A naive \emph{vev} for these charged states would result in brane recombination incompatible with local tadpole cancellation. It is reasonable to assume that such \emph{vev}'s do not represent extrema of the effective potential which is generated in these non supersymmetric vacua. Therefore, both in heterotic and orientifold models a control of the marginal deformations associated to twisted states or a complete knowledge of the scalar potential is essential in order to establish whether these deformations are truly marginal or not. This is an important and still open problem in the field which requires dedicated investigation, but this lies beyond the scope of this work.

Given the lack of supersymmetry, instabilities can occur at the loop level, induced by a modified geometry due to a non-vanishing dilaton tadpole on the world-sheet torus. This is indeed the case for the AdS vacua of \cite{Gubser:2001zr, Mourad:2016xbk} where tachyonic fluctuations develop around the new vacuum \cite{Basile:2018irz},  although a suitable interplay between tadpoles and  fluxes can recover a stable AdS vacuum  \cite{Baykara:2022cwj}\footnote{This type of problem is recurrent for the non-tachyonic ten-dimensional non-supersymmetric strings and for their Dudas-Mourad vacua \cite{Dudas:2000ff}, whose stability was analysed in further detail in \cite{Mourad:2023wjg}.}. We expect that these (in)stabilities will also be present in our constructions and they certainly deserve further study.


\vskip .2in
\section*{Acknowledgements}
Diego Perugini is grateful to the Centre de Physique Th\'eorique of \'Ecole Polytechnique and the 
Institut de Physique Th\'eorique of the Universit\'e Paris Saclay for the hospitality extended to him during the Erasmus+ stage.

\newpage

\bibliographystyle{utphys}

\begin{thebibliography}{10}


\bibitem{Antoniadis:1991kh}
I.~Antoniadis and C.~Kounnas, ``{Superstring phase transition at high
  temperature},'' \href{http://dx.doi.org/10.1016/0370-2693(91)90442-S}{{\em
  Phys. Lett. B} {\bfseries 261} (1991) 369--378}.

\bibitem{Antoniadis:1999gz}
I.~Antoniadis, J.~P. Derendinger, and C.~Kounnas, ``{Nonperturbative
  temperature instabilities in N=4 strings},''
  \href{http://dx.doi.org/10.1016/S0550-3213(99)00171-6}{{\em Nucl. Phys. B}
  {\bfseries 551} (1999) 41--77},
  \href{http://arxiv.org/abs/hep-th/9902032}{{\ttfamily arXiv:hep-th/9902032}}.

\bibitem{Kaidi:2020jla}
J.~Kaidi, ``{Stable Vacua for Tachyonic Strings},''
  \href{http://dx.doi.org/10.1103/PhysRevD.103.106026}{{\em Phys. Rev. D}
  {\bfseries 103} no.~10, (2021) 106026},
  \href{http://arxiv.org/abs/2010.10521}{{\ttfamily arXiv:2010.10521
  [hep-th]}}.

\bibitem{Hellerman:2004qa}
S.~Hellerman and X.~Liu, ``{Dynamical dimension change in supercritical string
  theory},'' \href{http://arxiv.org/abs/hep-th/0409071}{{\ttfamily
  arXiv:hep-th/0409071}}.

\bibitem{Ginsparg:1986wr}
P.~H. Ginsparg and C.~Vafa, ``{Toroidal Compactification of Nonsupersymmetric
  Heterotic Strings},''
  \href{http://dx.doi.org/10.1016/0550-3213(87)90387-7}{{\em Nucl. Phys. B}
  {\bfseries 289} (1987) 414}.

\bibitem{Fraiman:2023cpa}
B.~Fraiman, M.~Gra\~na, H.~Parra De~Freitas, and S.~Sethi,
  ``{Non-Supersymmetric Heterotic Strings on a Circle},''
  \href{http://arxiv.org/abs/2307.13745}{{\ttfamily arXiv:2307.13745
  [hep-th]}}.

\bibitem{Alvarez-Gaume:1986ghj}
L.~Alvarez-Gaume, P.~H. Ginsparg, G.~W. Moore, and C.~Vafa, ``{An $O(16) \times
  O(16)$ Heterotic String},''
  \href{http://dx.doi.org/10.1016/0370-2693(86)91524-8}{{\em Phys. Lett. B}
  {\bfseries 171} (1986) 155--162}.

\bibitem{Dixon:1986iz}
L.~J. Dixon and J.~A. Harvey, ``{String Theories in Ten-Dimensions Without
  Space-Time Supersymmetry},''
  \href{http://dx.doi.org/10.1016/0550-3213(86)90619-X}{{\em Nucl. Phys. B}
  {\bfseries 274} (1986) 93--105}.

\bibitem{Sagnotti:1995ga}
A.~Sagnotti, ``{Some properties of open string theories},'' in {\em
  {International Workshop on Supersymmetry and Unification of Fundamental
  Interactions (SUSY 95)}}, pp.~473--484.
\newblock 9, 1995.
\newblock \href{http://arxiv.org/abs/hep-th/9509080}{{\ttfamily
  arXiv:hep-th/9509080}}.

\bibitem{Sagnotti:1996qj}
A.~Sagnotti, ``{Surprises in open string perturbation theory},''
  \href{http://dx.doi.org/10.1016/S0920-5632(97)00344-7}{{\em Nucl. Phys. B
  Proc. Suppl.} {\bfseries 56} (1997) 332--343},
  \href{http://arxiv.org/abs/hep-th/9702093}{{\ttfamily arXiv:hep-th/9702093}}.

\bibitem{Sugimoto:1999tx}
S.~Sugimoto, ``{Anomaly cancellations in type I D-9 - anti-D-9 system and the
  USp(32) string theory},'' \href{http://dx.doi.org/10.1143/PTP.102.685}{{\em
  Prog. Theor. Phys.} {\bfseries 102} (1999) 685--699},
  \href{http://arxiv.org/abs/hep-th/9905159}{{\ttfamily arXiv:hep-th/9905159}}.

\bibitem{Antoniadis:1999xk}
I.~Antoniadis, E.~Dudas, and A.~Sagnotti, ``{Brane supersymmetry breaking},''
  \href{http://dx.doi.org/10.1016/S0370-2693(99)01023-0}{{\em Phys. Lett. B}
  {\bfseries 464} (1999) 38--45},
  \href{http://arxiv.org/abs/hep-th/9908023}{{\ttfamily arXiv:hep-th/9908023}}.

\bibitem{Aldazabal:1999jr}
G.~Aldazabal and A.~M. Uranga, ``{Tachyon free nonsupersymmetric type IIB
  orientifolds via Brane - anti-brane systems},''
  \href{http://dx.doi.org/10.1088/1126-6708/1999/10/024}{{\em JHEP} {\bfseries
  10} (1999) 024}, \href{http://arxiv.org/abs/hep-th/9908072}{{\ttfamily
  arXiv:hep-th/9908072}}.

\bibitem{Angelantonj:1999jh}
C.~Angelantonj, ``{Comments on open string orbifolds with a nonvanishing
  $B_{ab}$},'' \href{http://dx.doi.org/10.1016/S0550-3213(99)00662-8}{{\em
  Nucl. Phys. B} {\bfseries 566} (2000) 126--150},
  \href{http://arxiv.org/abs/hep-th/9908064}{{\ttfamily arXiv:hep-th/9908064}}.

\bibitem{Angelantonj:1999ms}
C.~Angelantonj, I.~Antoniadis, G.~D'Appollonio, E.~Dudas, and A.~Sagnotti,
  ``{Type I vacua with brane supersymmetry breaking},''
  \href{http://dx.doi.org/10.1016/S0550-3213(00)00052-3}{{\em Nucl. Phys. B}
  {\bfseries 572} (2000) 36--70},
  \href{http://arxiv.org/abs/hep-th/9911081}{{\ttfamily arXiv:hep-th/9911081}}.

\bibitem{Angelantonj:2024iwi}
C.~Angelantonj, C.~Condeescu, E.~Dudas, and G.~Leone, ``{Rigid vacua with Brane
  Supersymmetry Breaking},''
  \href{http://dx.doi.org/10.1007/JHEP04(2024)103}{{\em JHEP} {\bfseries 04}
  (2024) 103}, \href{http://arxiv.org/abs/2403.02392}{{\ttfamily
  arXiv:2403.02392 [hep-th]}}.

\bibitem{Angelantonj:2006ut}
C.~Angelantonj, M.~Cardella, and N.~Irges, ``{An Alternative for Moduli
  Stabilisation},''
  \href{http://dx.doi.org/10.1016/j.physletb.2006.08.072}{{\em Phys. Lett. B}
  {\bfseries 641} (2006) 474--480},
  \href{http://arxiv.org/abs/hep-th/0608022}{{\ttfamily arXiv:hep-th/0608022}}.

\bibitem{Baykara:2024tjr}
Z.~K. Baykara, H.-C. Tarazi, and C.~Vafa, ``{New Non-Supersymmetric
  Tachyon-Free Strings},'' \href{http://arxiv.org/abs/2406.00185}{{\ttfamily
  arXiv:2406.00185 [hep-th]}}.

\bibitem{Harvey:1987da}
J.~A. Harvey, G.~W. Moore, and C.~Vafa, ``{Quasicrystalline
  Compactification},''
  \href{http://dx.doi.org/10.1016/0550-3213(88)90627-X}{{\em Nucl. Phys. B}
  {\bfseries 304} (1988) 269--290}.

\bibitem{Baykara:2024vss}
Z.~K. Baykara, H.-C. Tarazi, and C.~Vafa, ``{The Quasicrystalline String
  Landscape},'' \href{http://arxiv.org/abs/2406.00129}{{\ttfamily
  arXiv:2406.00129 [hep-th]}}.

\bibitem{Font:2021uyw}
A.~Font, B.~Fraiman, M.~Gra\~na, C.~A. N\'u\~nez, and H.~Parra De~Freitas,
  ``{Exploring the landscape of CHL strings on T$^{d}$},''
  \href{http://dx.doi.org/10.1007/JHEP08(2021)095}{{\em JHEP} {\bfseries 08}
  (2021) 095}, \href{http://arxiv.org/abs/2104.07131}{{\ttfamily
  arXiv:2104.07131 [hep-th]}}.

\bibitem{Nakajima:2023zsh}
S.~Nakajima, ``{New non-supersymmetric heterotic string theory with reduced
  rank and exponential suppression of the cosmological constant},''
  \href{http://arxiv.org/abs/2303.04489}{{\ttfamily arXiv:2303.04489
  [hep-th]}}.

\bibitem{DeFreitas:2024ztt}
H.~P. De~Freitas, ``{Non-supersymmetric heterotic strings and chiral CFTs},''
  \href{http://arxiv.org/abs/2402.15562}{{\ttfamily arXiv:2402.15562
  [hep-th]}}.

\bibitem{Chaudhuri:1995fk}
S.~Chaudhuri, G.~Hockney, and J.~D. Lykken, ``{Maximally supersymmetric string
  theories in D \ensuremath{<} 10},''
  \href{http://dx.doi.org/10.1103/PhysRevLett.75.2264}{{\em Phys. Rev. Lett.}
  {\bfseries 75} (1995) 2264--2267},
  \href{http://arxiv.org/abs/hep-th/9505054}{{\ttfamily arXiv:hep-th/9505054}}.

\bibitem{Chaudhuri:1995bf}
S.~Chaudhuri and J.~Polchinski, ``{Moduli space of CHL strings},''
  \href{http://dx.doi.org/10.1103/PhysRevD.52.7168}{{\em Phys. Rev. D}
  {\bfseries 52} (1995) 7168--7173},
  \href{http://arxiv.org/abs/hep-th/9506048}{{\ttfamily arXiv:hep-th/9506048}}.

\bibitem{Narain:1986qm}
K.~S. Narain, M.~H. Sarmadi, and C.~Vafa, ``{Asymmetric Orbifolds},''
  \href{http://dx.doi.org/10.1016/0550-3213(87)90228-8}{{\em Nucl. Phys. B}
  {\bfseries 288} (1987) 551}.

\bibitem{Narain:1990mw}
K.~S. Narain, M.~H. Sarmadi, and C.~Vafa, ``{Asymmetric orbifolds: Path
  integral and operator formulations},''
  \href{http://dx.doi.org/10.1016/0550-3213(91)90145-N}{{\em Nucl. Phys. B}
  {\bfseries 356} (1991) 163--207}.

\bibitem{Ibanez:1987pj}
L.~E. Ibanez, J.~Mas, H.-P. Nilles, and F.~Quevedo, ``{Heterotic Strings in
  Symmetric and Asymmetric Orbifold Backgrounds},''
  \href{http://dx.doi.org/10.1016/0550-3213(88)90166-6}{{\em Nucl. Phys. B}
  {\bfseries 301} (1988) 157--196}.

\bibitem{Beye:2013ola}
F.~Beye, T.~Kobayashi, and S.~Kuwakino, ``{Three-generation Asymmetric Orbifold
  Models from Heterotic String Theory},''
  \href{http://dx.doi.org/10.1007/JHEP01(2014)013}{{\em JHEP} {\bfseries 01}
  (2014) 013}, \href{http://arxiv.org/abs/1311.4687}{{\ttfamily arXiv:1311.4687
  [hep-th]}}.

\bibitem{Acharya:2022shu}
B.~S. Acharya, G.~Aldazabal, A.~Font, K.~Narain, and I.~G. Zadeh, ``{Heterotic
  strings on $ T^3/\mathbb{Z}_{2}$, Nikulin involutions and M-theory},''
  \href{http://dx.doi.org/10.1007/JHEP09(2022)209}{{\em JHEP} {\bfseries 09}
  (2022) 209}, \href{http://arxiv.org/abs/2205.09764}{{\ttfamily
  arXiv:2205.09764 [hep-th]}}.

\bibitem{Gkountoumis:2023fym}
G.~Gkountoumis, C.~Hull, K.~Stemerdink, and S.~Vandoren, ``{Freely acting
  orbifolds of type IIB string theory on $T^{5}$},''
  \href{http://dx.doi.org/10.1007/JHEP08(2023)089}{{\em JHEP} {\bfseries 08}
  (2023) 089}, \href{http://arxiv.org/abs/2302.09112}{{\ttfamily
  arXiv:2302.09112 [hep-th]}}.

\bibitem{Gkountoumis:2024dwc}
G.~Gkountoumis, C.~Hull, and S.~Vandoren, ``{Exact moduli spaces for
  $\mathcal{N}=2$, $D=5$ freely acting orbifolds},''
  \href{http://arxiv.org/abs/2403.05650}{{\ttfamily arXiv:2403.05650
  [hep-th]}}.

\bibitem{Bianchi:2012xz}
M.~Bianchi, G.~Pradisi, C.~Timirgaziu, and L.~Tripodi, ``{Heterotic T-folds
  with a small number of neutral moduli},''
  \href{http://dx.doi.org/10.1007/JHEP10(2012)089}{{\em JHEP} {\bfseries 10}
  (2012) 089}, \href{http://arxiv.org/abs/1207.2665}{{\ttfamily arXiv:1207.2665
  [hep-th]}}.

\bibitem{Anastasopoulos:2009kj}
P.~Anastasopoulos, M.~Bianchi, J.~F. Morales, and G.~Pradisi, ``{(Unoriented)
  T-folds with few T's},''
  \href{http://dx.doi.org/10.1088/1126-6708/2009/06/032}{{\em JHEP} {\bfseries
  06} (2009) 032}, \href{http://arxiv.org/abs/0901.0113}{{\ttfamily
  arXiv:0901.0113 [hep-th]}}.

\bibitem{Bianchi:1999uq}
M.~Bianchi, J.~F. Morales, and G.~Pradisi, ``{Discrete torsion in nongeometric
  orbifolds and their open string descendants},''
  \href{http://dx.doi.org/10.1016/S0550-3213(99)00765-8}{{\em Nucl. Phys. B}
  {\bfseries 573} (2000) 314--334},
  \href{http://arxiv.org/abs/hep-th/9910228}{{\ttfamily arXiv:hep-th/9910228}}.

\bibitem{Baykara:2023plc}
Z.~K. Baykara, Y.~Hamada, H.-C. Tarazi, and C.~Vafa, ``{On the string landscape
  without hypermultiplets},''
  \href{http://dx.doi.org/10.1007/JHEP04(2024)121}{{\em JHEP} {\bfseries 04}
  (2024) 121}, \href{http://arxiv.org/abs/2309.15152}{{\ttfamily
  arXiv:2309.15152 [hep-th]}}.

\bibitem{Condeescu:2012sp}
C.~Condeescu, I.~Florakis, and D.~Lust, ``{Asymmetric Orbifolds, Non-Geometric
  Fluxes and Non-Commutativity in Closed String Theory},''
  \href{http://dx.doi.org/10.1007/JHEP04(2012)121}{{\em JHEP} {\bfseries 04}
  (2012) 121}, \href{http://arxiv.org/abs/1202.6366}{{\ttfamily arXiv:1202.6366
  [hep-th]}}.

\bibitem{Condeescu:2013yma}
C.~Condeescu, I.~Florakis, C.~Kounnas, and D.~L\"ust, ``{Gauged supergravities
  and non-geometric Q/R-fluxes from asymmetric orbifold CFT`s},''
  \href{http://dx.doi.org/10.1007/JHEP10(2013)057}{{\em JHEP} {\bfseries 10}
  (2013) 057}, \href{http://arxiv.org/abs/1307.0999}{{\ttfamily arXiv:1307.0999
  [hep-th]}}.

\bibitem{Angelantonj:2002ct}
C.~Angelantonj and A.~Sagnotti, ``{Open strings},''
  \href{http://dx.doi.org/10.1016/S0370-1573(02)00273-9}{{\em Phys. Rept.}
  {\bfseries 371} (2002) 1--150},
  \href{http://arxiv.org/abs/hep-th/0204089}{{\ttfamily arXiv:hep-th/0204089}}.
  [Erratum: Phys.Rept. 376, 407 (2003)].

\bibitem{Green:1984sg}
M.~B. Green and J.~H. Schwarz, ``{Anomaly Cancellation in Supersymmetric $D=10$
  Gauge Theory and Superstring Theory},''
  \href{http://dx.doi.org/10.1016/0370-2693(84)91565-X}{{\em Phys. Lett. B}
  {\bfseries 149} (1984) 117--122}.

\bibitem{Sagnotti:1992qw}
A.~Sagnotti, ``{A Note on the Green-Schwarz mechanism in open string
  theories},'' \href{http://dx.doi.org/10.1016/0370-2693(92)90682-T}{{\em Phys.
  Lett. B} {\bfseries 294} (1992) 196--203},
  \href{http://arxiv.org/abs/hep-th/9210127}{{\ttfamily arXiv:hep-th/9210127}}.

\bibitem{Schellekens:1986yi}
A.~N. Schellekens and N.~P. Warner, ``{Anomalies and Modular Invariance in
  String Theory},'' \href{http://dx.doi.org/10.1016/0370-2693(86)90760-4}{{\em
  Phys. Lett. B} {\bfseries 177} (1986) 317--323}.

\bibitem{Schellekens:1986xh}
A.~N. Schellekens and N.~P. Warner, ``{Anomalies, Characters and Strings},''
  \href{http://dx.doi.org/10.1016/0550-3213(87)90108-8}{{\em Nucl. Phys. B}
  {\bfseries 287} (1987) 317}.

\bibitem{Basile:2023zng}
I.~Basile and G.~Leone, ``{Anomaly constraints for heterotic strings and
  supergravity in six dimensions},''
  \href{http://dx.doi.org/10.1007/JHEP04(2024)067}{{\em JHEP} {\bfseries 04}
  (2024) 067}, \href{http://arxiv.org/abs/2310.20480}{{\ttfamily
  arXiv:2310.20480 [hep-th]}}.

\bibitem{Kutasov:1990sv}
D.~Kutasov and N.~Seiberg, ``{Number of degrees of freedom, density of states
  and tachyons in string theory and CFT},''
  \href{http://dx.doi.org/10.1016/0550-3213(91)90426-X}{{\em Nucl. Phys. B}
  {\bfseries 358} (1991) 600--618}.

\bibitem{Dienes:1994np}
K.~R. Dienes, ``{Modular invariance, finiteness, and misaligned supersymmetry:
  New constraints on the numbers of physical string states},''
  \href{http://dx.doi.org/10.1016/0550-3213(94)90153-8}{{\em Nucl. Phys. B}
  {\bfseries 429} (1994) 533--588},
  \href{http://arxiv.org/abs/hep-th/9402006}{{\ttfamily arXiv:hep-th/9402006}}.

\bibitem{Angelantonj:2010ic}
C.~Angelantonj, M.~Cardella, S.~Elitzur, and E.~Rabinovici, ``{Vacuum
  stability, string density of states and the Riemann zeta function},''
  \href{http://dx.doi.org/10.1007/JHEP02(2011)024}{{\em JHEP} {\bfseries 02}
  (2011) 024}, \href{http://arxiv.org/abs/1012.5091}{{\ttfamily arXiv:1012.5091
  [hep-th]}}.

\bibitem{Cribiori:2020sct}
N.~Cribiori, S.~Parameswaran, F.~Tonioni, and T.~Wrase, ``{Misaligned
  Supersymmetry and Open Strings},''
  \href{http://dx.doi.org/10.1007/JHEP04(2021)099}{{\em JHEP} {\bfseries 04}
  (2021) 099}, \href{http://arxiv.org/abs/2012.04677}{{\ttfamily
  arXiv:2012.04677 [hep-th]}}.

\bibitem{Angelantonj:2023egh}
C.~Angelantonj, I.~Florakis, and G.~Leone, ``{Tachyons and misaligned
  supersymmetry in closed string vacua},''
  \href{http://dx.doi.org/10.1007/JHEP06(2023)174}{{\em JHEP} {\bfseries 06}
  (2023) 174}, \href{http://arxiv.org/abs/2301.13702}{{\ttfamily
  arXiv:2301.13702 [hep-th]}}.

\bibitem{Leone:2023qfd}
G.~Leone, ``{Tachyons and Misaligned Supersymmetry in orientifold vacua},''
  \href{http://dx.doi.org/10.1007/JHEP11(2023)066}{{\em JHEP} {\bfseries 11}
  (2023) 066}, \href{http://arxiv.org/abs/2308.09757}{{\ttfamily
  arXiv:2308.09757 [hep-th]}}.

\bibitem{Antoniadis:1986rn}
I.~Antoniadis, C.~P. Bachas, and C.~Kounnas, ``{Four-Dimensional
  Superstrings},'' \href{http://dx.doi.org/10.1016/0550-3213(87)90372-5}{{\em
  Nucl. Phys. B} {\bfseries 289} (1987) 87}.

\bibitem{Florakis:2024ubz}
I.~Florakis and J.~I. Rizos, {\em {Free Fermionic Constructions of Heterotic
  Strings}}.
\newblock 7, 2024.
\newblock \href{http://arxiv.org/abs/2407.07034}{{\ttfamily arXiv:2407.07034
  [hep-th]}}.

\bibitem{Vafa:1994rv}
C.~Vafa and E.~Witten, ``{On orbifolds with discrete torsion},''
  \href{http://dx.doi.org/10.1016/0393-0440(94)00048-9}{{\em J. Geom. Phys.}
  {\bfseries 15} (1995) 189--214},
  \href{http://arxiv.org/abs/hep-th/9409188}{{\ttfamily arXiv:hep-th/9409188}}.

\bibitem{Gubser:2001zr}
S.~S. Gubser and I.~Mitra, ``{Some interesting violations of the
  Breitenlohner-Freedman bound},''
  \href{http://dx.doi.org/10.1088/1126-6708/2002/07/044}{{\em JHEP} {\bfseries
  07} (2002) 044}, \href{http://arxiv.org/abs/hep-th/0108239}{{\ttfamily
  arXiv:hep-th/0108239}}.

\bibitem{Mourad:2016xbk}
J.~Mourad and A.~Sagnotti, ``{$AdS$ Vacua from Dilaton Tadpoles and Form
  Fluxes},'' \href{http://dx.doi.org/10.1016/j.physletb.2017.02.053}{{\em Phys.
  Lett. B} {\bfseries 768} (2017) 92--96},
  \href{http://arxiv.org/abs/1612.08566}{{\ttfamily arXiv:1612.08566
  [hep-th]}}.

\bibitem{Basile:2018irz}
I.~Basile, J.~Mourad, and A.~Sagnotti, ``{On Classical Stability with Broken
  Supersymmetry},'' \href{http://dx.doi.org/10.1007/JHEP01(2019)174}{{\em JHEP}
  {\bfseries 01} (2019) 174}, \href{http://arxiv.org/abs/1811.11448}{{\ttfamily
  arXiv:1811.11448 [hep-th]}}.

\bibitem{Baykara:2022cwj}
Z.~K. Baykara, D.~Robbins, and S.~Sethi, ``{Non-supersymmetric AdS from string
  theory},'' \href{http://dx.doi.org/10.21468/SciPostPhys.15.6.224}{{\em
  SciPost Phys.} {\bfseries 15} no.~6, (2023) 224},
  \href{http://arxiv.org/abs/2212.02557}{{\ttfamily arXiv:2212.02557
  [hep-th]}}.

\bibitem{Dudas:2000ff}
E.~Dudas and J.~Mourad, ``{Brane solutions in strings with broken supersymmetry
  and dilaton tadpoles},''
  \href{http://dx.doi.org/10.1016/S0370-2693(00)00734-6}{{\em Phys. Lett. B}
  {\bfseries 486} (2000) 172--178},
  \href{http://arxiv.org/abs/hep-th/0004165}{{\ttfamily arXiv:hep-th/0004165}}.

\bibitem{Mourad:2023wjg}
J.~Mourad and A.~Sagnotti, ``{Non-supersymmetric vacua and self-adjoint
  extensions},'' \href{http://dx.doi.org/10.1007/JHEP08(2023)041}{{\em JHEP}
  {\bfseries 08} (2023) 041}, \href{http://arxiv.org/abs/2305.09587}{{\ttfamily
  arXiv:2305.09587 [hep-th]}}.


\end{thebibliography}

\end{document}